\documentclass{jfm}
\usepackage{graphicx}
\usepackage{epstopdf, epsfig}
\usepackage{amssymb, amsmath, color, eurosym, natbib}

\newcommand{\bA}{\mathbf{A}}
\newcommand{\bB}{\mathbf{B}}
\newcommand{\bF}{\mathbf{F}}
\newcommand{\bI}{\mathbf{I}}
\newcommand{\bU}{\mathbf{U}}

\newcommand{\bz}{\mathbf{z}}

\shorttitle{Internal waves and mixing in a three-layer flow}
\shortauthor{A. A. Chesnokov, S. L. Gavrilyuk and V. Yu. Liapidevskii}
\title{Long nonlinear internal waves and mixing in a three-layer stratified flow in the Boussinesq approximation}
\author{A. A. Chesnokov\aff{1}
	\corresp{\email{chesnokov@hydro.nsc.ru}},
	S. L. Gavrilyuk\aff{2}
	\and V. Yu. Liapidevskii\aff{1}}

\affiliation{ 
	\aff{1}Lavrentyev Institute of Hydrodynamics, 15 Lavrentyev prospect, 630090 Novosibirsk, Russia 
	\aff{2}Aix-Marseille Universit\'{e}, UMR CNRS 7343, IUSTI, 5 rue E. Fermi, 13453 Marseille CEDEX 13, France}

\begin{document}
\maketitle

\begin{abstract}
A one-dimensional long-wave model of an unsteady three-layer flow of a stratified fluid under a lid is proposed, taking into account  turbulent mixing in the intermediate layer. In the Boussinesq approximation, the equations of motion are reduced to an evolutionary system of balance laws, which is hyperbolic for a small difference in velocities in the layers. Classes of stationary solutions are studied and the concept of subcritical (supercritical) three-layer stratified flow is introduced. Oscillating solutions are constructed that describe the spatial evolution of the mixing layer. The problem of transcritical flow over an obstacle is considered. Solutions are obtained that describe qualitatively different flow regimes on the leeward side of the obstacle. The proposed model is validated using  experimental data and field observations on the entrainment of ambient fluid. 
\end{abstract}
		

\section{Introduction}
\label{sec:intro}

Stratified flows over topography are ubiquitous in geophysical and environmental settings. For instance, the flow of a dense fluid along a sloping bottom occurs  in  deep ocean outflows and severe downslope winds. A detailed overview  of the diversity of such flows and their relevance can be found  in the monographs of \cite{Baines05} and  \cite{Simpson97}.

A distinguishing feature of stratified flow over topography is the formation of a bifurcation enclosing partly mixed fluid. Detailed observations of the establishment of such topographic flows over the Knight Inlet sill  (British Columbia, Canada) are  described in  \cite{Farmer99, Armi02, Cummins06}. In particular, in the last reference, the upstream formation of a strong internal hydraulic jump is reported and studied. Similar phenomena are reproduced in the laboratory experiments of \cite{Pawlak98, Pawlak00} where the special character of shear instability in steep downslope flows is illustrated. The enhanced entrainment efficiency due to  the Kelvin--Helmholtz instability leads to the changing density of the trapped fluid, thus providing the link between small-scale processes and the larger-scale response. A Navier--Stokes type model of stratified flows is used in  \cite{Lamb04} to describe the formation of a strong supercritical flow beneath a large breaking lee wave in tidal flow over the Knight Inlet sill. Further study of topographic effects in stratified flows is carried out \cite{Winters14}. A set of example flows is introduced that systematically leads to the observation that approximately blocked and topographically controlled flows are produced whenever a uniform inflow encounters an obstacle that is sufficiently tall. The dynamic stability and connection between topographic control and wave excitation aloft of stratified flow configurations characteristic of hydraulically controlled downslope flow over topography are studied in recent works of \cite{Jagannathan17, Jagannathan20}.

In theoretical modelling of turbulent shear flows, such as  a surface or near-bottom jet, or  a mixing layer, an important question is how turbulence is generated and how it maintains itself in a stably gravity-stratified shear flow. \cite{Chu84} considered the characteristic features of the stratification influence on mixing processes in the near-field and far-field. \cite{Sher15, Sher17} performed experiments of a turbulent gravity current and presented  measurements of the entrainment of ambient fluid into gravity currents produced by a steady flux of buoyancy. It is known that the process of entrainment produces a deepening mixing layer at the interface, which increases the gradient Richardson number of this layer and eventually may suppress further entrainment. Taking this fact into account, \cite{Horsley18} constructed an analytical solution to the simplified depth-averaged model and discussed its properties. \cite{Yuan17} presented experimental investigation of large-scale vortices in a freely spreading gravity currents propagating into laterally confined and unconfined environments. 

Internal turbulent hydraulic jump is another  possible mechanisms of mixing.  \cite{Ogden16} investigated internal hydraulic jumps in flows with upstream shear using two-layer shock-joining theories. The models can be modified to allow entrainment and to indirectly account for continuous velocity profiles, producing solutions where the basic theories failed. \cite{Ogden20} studied these jumps to illuminate the changing physics as shear increases. \cite{Baines16} presented a two-layer model of internal hydraulic jumps which incorporates mixing between the layers within the jump. \cite{Thorpe14, Thorpe18} derived another model of internal hydraulic jumps which adopts continuous profiles of velocity and density both upstream and downstream of the transition. They also applied this model to describe observations made at several locations in the ocean.

The  aim of the present  paper is  to propose a mathematical model for the formation and evolution of a mixing layer resulting from  the  interaction between two co-flowing layers of homogeneous fluids over topography. Experimental data and field observations on gravity currents presented in the above-mentioned works indicate the presence of a fairly clear boundaries between the regions of potential flow of homogeneous outer layers with different densities and an intermediate non-homogeneous layer of turbulent mixing. The outer layers are potential and can be approximately described by a shallow water type model. In the intermediate layer, the flow is sheared and is described by the equations of weakly sheared flows (\cite{Tesh07}). The interaction between the outer layers and interlayer is taken into account through a natural mixing process, where the mixing velocity is proportional to the intensity of large eddies in the interlayer. This approach has already been used to simulate a mixing layer in a homogeneous fluid (\cite{LiapChesn14, ChesnLiap20}) and to describe the evolution of spilling breakers in shallow water (\cite{GLCh16}). Its extension to the case of inhomogeneous stratified flows is proposed by \cite{GLCh19}. A similar approach based on the layered description of the stratified flow is used by \cite{Liap04, LiapDG18} to simulate the mixing layer on the leeward side of the obstacle and sediment laden gravity currents. However, in the  last references, the process of entrainment of liquid into the mixing layer is taken into account in a slightly different way. 

In the present work,  we assume that the upper boundary is fixed and the density difference in the layers is small, so we can use the Boussinesq approximation. This makes it possible to obtain a depth-averaged model of a three-layer system consisting of two outer homogeneous layers  and an intermediate mixing layer in the form of a system of six balance laws. The system is  hyperbolic for  small relative  velocities  in the layers. This model allows for a simple numerical implementation and is used to describe the basic  mixing features in stratified flows over a flat and uneven bottom.

The three-layer flow scheme makes it possible to eliminate contradictions that cannot be resolved in the framework of a two-layer scheme. Among them a problem of choosing relations between the downstream and upstream states of an internal hydraulic jump, as well as the need to take into account the non-uniformity of  velocity profile and mass transfer between the layers by additional empirical relations for hydraulic jumps (\cite{Ogden16}). The model derived  in this paper is a  closed system of integral conservation laws describing a three-layer shallow water flow over topography. The model allows one to describe turbulent mixing between homogeneous layers as a nonlinear stage of the Kelvin--Helmholtz instability development at the interface between layers both in stationary and non-stationary  flows. The non-uniform  velocity profile (long-wave horizontal vorticity) in the mixing layer  is taken into account. The main attention in the work is focused on the study of the capabilities of the presented model for the mathematical description of the  effects of mixing and topography on the structure of supercritical and subcritical flows of stratified fluids. Applications of this model for the quantitative description of laboratory experiments and field observations are presented to show its ability to describe complex stratified flows. 

In the following section, a one-dimensional long-wave model of a three-layer stratified flow under a lid is derived. The model takes into account turbulent mixing between adjacent layers. In Section \ref{sec:Bouss_model},  the Boussinesq approximation is performed and the characteristic velocities are determined. Stationary solutions are studied in Section \ref{sec:stationary}, where the definition of subcritical (supercritical) three-layer flows is proposed. Then, continuous and discontinuous oscillating solutions are constructed that describe the spatial evolution of the mixing layer. In Section \ref{sec:model_validation}, the model is validated  by comparison with known experimental data and field observations.  Finally, conclusions are drawn in Section \ref{sec:conclusion}.

\section{Long-wave model of a three-layer flow with mixing} 
\label{sec:general_model}

\begin{figure}
	\begin{center}
		\resizebox{0.5\textwidth}{!}{\includegraphics{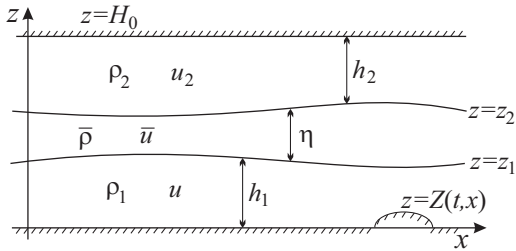}}\\[0pt]
		{\caption{Three-layer stratified flow over topography.} \label{fig:fig_1}} 
	\end{center}
\end{figure}

Consider  two-dimensional non-stationary flows  of an inviscid incompressible non-homogeneous fluid confined between the moving bottom $z=Z(t,x)$ and rigid upper lid $z=H_0={\rm const}$. The corresponding dimensionless Euler equations are 
\begin{equation} \label{eq:Euler}  
 \begin{array}{l}\displaystyle
  u_x+w_z=0, \quad \rho_t+ (u\rho)_x+ (w\rho)_z=0, \\[3mm]\displaystyle
  (\rho u)_t+ (\rho u^2)_x+ (\rho uw)_z+ p_x=0, \\[3mm]\displaystyle
  \varepsilon^2\big((\rho w)_t+ (\rho uw)_x+ (\rho w^2)_z\big)+ p_z=-g\rho.
 \end{array}
\end{equation} 
The kinematic boundary conditions at $z=Z$ and $z=H_0$ are
\begin{equation} \label{eq:BC} 
  Z_t+uZ_x-w\big|_{z=Z}=0, \quad w\big|_{z=H_0}=0. 
\end{equation} 
In what follows, we assume that the flow has a three-layer structure (Fig.~\ref{fig:fig_1}). The internal boundaries $z=z_1=Z+h_1(t,x)$ and $z=z_2=H_0-h_2(t,x)$ separate the outer layers, where the flow is almost potential  and homogeneous with constant densities $\rho_1$ and $\rho_2$, and the intermediate turbulent non-homogeneous layer with density $\rho(t,x,z)$. The vertical density distribution is assumed to be continuous so that 
\begin{equation} \label{eq:BC-rho} 
  \rho\big|_{z=z_1+0}=\rho_1, \quad \rho\big|_{z=z_2-0}=\rho_2 .
\end{equation} 
The  depth-averaged density of the intermediate layer is defined as  
 \[ \bar{\rho}(t,x) =\frac{1}{\eta}\int_{z_1}^{z_2}\rho(t,x,z')dz'.\] 
We will show later that the vertical density distribution  will be stable for any time, i.e. $\rho_2<\bar{\rho}(t,x)<\rho_1$, if it was initially stable. 
At the internal boundaries  the kinematic conditions are satisfied:
\begin{equation} \label{eq:BC-interfaces}  
	z_{1t}+uz_{1x}-w\big|_{z=z_1}=-M_1, \quad z_{2t}+uz_{2x}-w\big|_{z=z_2}=M_2. 
\end{equation} 
The right-hand sides  $M_1$ and $M_2$ responsible for the mixing between layers will be precised  later. In the above equations $u=V^{-1}\hat{u}$, $w=(d_0 V)^{-1}l_0 \hat{w}$, $p=(\rho_0 V^2)^{-1}\hat{p}$, $x=l_0^{-1}\hat{x}$, $z=d_0^{-1}\hat{z}$, $t=Vl_0^{-1}\hat{t}$, and $g=d_0 V^{-2}\hat{g}$ are dimensionless components of the velocity, pressure, Cartesian coordinates, time, and gravitational acceleration, respectively; $\hat{u}$, $\hat{w}$, $\hat{p}$, $\hat{x}$, $\hat{z}$, $\hat{t}$, and $\hat{g}$ are the corresponding dimensional variables. The parameters $V$, $\rho_0$, $d_0$, $l_0$ denote the characteristic velocity, density, and the characteristic vertical and horizontal scales, respectively. The dimensionless parameter $\varepsilon=d_0/h_0$ is the ratio of vertical and horizontal scales. System (\ref{eq:Euler}) admits the conservation of energy 
\begin{equation} \label{eq:energy} 
E_t+ \big((E+p)u\big)_x+ \big((E+p)w\big)_z=0,
\end{equation} 
where $E=(u^2/2+\varepsilon^2 w^2/2+gz)\rho$. We suppose that the waves are long, i.e. $\varepsilon\ll 1$ and the terms of order $O(\varepsilon^2)$ are neglected in equations (\ref{eq:Euler}), (\ref{eq:energy}). In this case, the last equation in system (\ref{eq:Euler}) is reduced to the hydrostatic law of pressure distribution over the depth
\begin{equation} \label{eq:pressure-gen}  
p=p^*(t,x)+g\int\limits_z^{H_0}\rho(t,x,z')\,dz', 
\end{equation}
where $p^*$ is the pressure at the upper lid. 

To describe the three-layer flow, we define the following depth-averaged variables:
\begin{equation} \label{eq:variables} 
 \begin{array}{l}\displaystyle 
  u_1=\frac{1}{h_1}\int\limits_Z^{z_1} u\,dz, \quad u_2=\frac{1}{h_2} \int\limits_{z_2}^{H_0} u\,dz, \quad \bar{u}=\frac{1}{\eta}\int\limits_{z_1}^{z_2} u\,dz, \\[3mm]\displaystyle
  \quad\quad\quad
  q^2=\frac{1}{\eta}\int\limits_{z_1}^{z_2} (u-\bar{u})^2\,dz, \quad 
  \bar{\rho}=\frac{1}{\eta}\int\limits_{z_1}^{z_2} \rho\,dz.  
 \end{array} 
\end{equation} 
Here $\eta=z_2-z_1$ is the  thickness of the intermediate layer. The variable $q$ measures the distortion of the velocity profile in the mixing layer. Note that the geometric constraint yields $h_1+\eta+h_2=H_0-Z$. Further we derive a one-dimensional closed system of depth-averaged equations.

\subsection{Depth-averaged equations for the outer layers}

As mentioned above, in the lower and upper layers the fluid is homogeneous ($\rho_i={\rm const}$, $i=1,2$).  From~(\ref{eq:pressure-gen}) it follows that the pressure in these layers is 
\begin{equation}\label{eq:pressure-out}
  p= \left\{
    \begin{array}{ll}
      (z_1-z)g\rho_1+g\bar{\rho}\eta+g\rho_2 h_2+p^*,   & \ z \in (Z, z_1)  \\[2mm]
      (H_0-z)g\rho_2+p^*  & \ z \in (z_2, H_0)
    \end{array} \right. 
\end{equation}
The derivation of  averaged over depth equations for outer layers is similar.  For definiteness, we will derive in details the equations for  the upper layer.  Integrating the incompressibility, horizontal momentum and energy equations with respect to $z$ over the interval $(z_2, H_0)$ and using the boundary conditions, we get the following exact integral relations:
\begin{equation} \label{eq:upper-layer-1} 
 \begin{array}{l}\displaystyle 
   h_{2t}+(u_2 h_2)_x=-M_2, \\[3mm]\displaystyle
  (u_2h_2)_t+ \bigg(\int\limits_{z_2}^{H_0} \Big(u^2+\frac{p}{\rho_2}\Big)\,dz\bigg)_x= -\Big(M_2 u+\frac{p}{\rho_2}z_{2x}\Big)\Big|_{z=z_2+0}\,, \\[3mm]\displaystyle
  \bigg(\int\limits_{z_2}^{H_0} E\,dz\bigg)_t+ \bigg(\int\limits_{z_2}^{H_0} (E+p)u\,dz\bigg)_x= -\big(M_2 (E+p)+h_{2t}p\big)\big|_{z=z_2+0}\,.
 \end{array} 
\end{equation}

The calculation of the integrals in system (\ref{eq:upper-layer-1}) is based on the estimates given below. We will say that the flow is {\it weakly sheared} if $u_z=O(\varepsilon^\alpha)$, $\alpha>0$. For a homogeneous fluid the flow vorticity $\omega=u_z-\varepsilon^2 w_x$ conserves along the trajectories: $\omega_t+u\omega_x+w\omega_z=0$. The long-wave vorticity $\omega=u_z$ also satisfies this equation with error $O(\varepsilon^2)$. Therefore, if $u_z=O(\varepsilon^\beta)$ for $t=0$, then $u_z=O(\varepsilon^\beta)$  for any $t>0$, with $\beta=\min(\alpha, 2)$. One can prove that, if the flow is weakly sheared, then 
\begin{equation}\label{eq:estimates-1} 
  \frac{1}{h_1}\int\limits_Z^{z_1} u^2\,dz=u_1^2+O(\varepsilon^{2\beta}), \quad   
  \frac{1}{h_2}\int\limits_{z_2}^{H_0} u^3\,dz=u_2^3+O(\varepsilon^{3 \beta}). 
\end{equation}
(for the proof, see \cite{Barros07, GLCh19}). 

We say that the  fluid flow in the outer layers is almost potential, if $\alpha >1$.   Since terms of order $O(\varepsilon^2)$ and higher are not taken into account in the considered long-wave approximation, this means the terms $O(\varepsilon^{2\beta})$ and $O(\varepsilon^{3\beta})$ in formulae~(\ref{eq:estimates-1}) are ignored. Neglecting higher order terms  system (\ref{eq:upper-layer-1})  takes the form
\begin{equation} \label{eq:upper-layer} 
 \begin{array}{l}\displaystyle 
  h_{2t}+(u_2 h_2)_x=-M_2, \\[3mm]\displaystyle
  (u_2h_2)_t+\Big(u_2^2h_2+ \frac{gh_2^2}{2}+ \frac{p^* h_2}{\rho_2} \Big)_x= -M_2 u\big|_{z=z_2+0}+ 
  \Big(\frac{p^*}{\rho_2}+gh_2\Big)h_{2x} \,, \\[3mm]\displaystyle
  \bigg(\frac{u_2^2 h_2}{2}+gH_0 h_2-\frac{gh_2^2}{2}\bigg)_t+ \bigg(\frac{u_2^3 h_2}{2}+ \Big(\frac{p^*}{\rho_2} + gH_0\Big)u_2 h_2\bigg)_x \\[3mm]\displaystyle
  \quad\quad =-\bigg(\frac{u^2}{2}\bigg|_{z=z_2+0} +gH_0+\frac{p^*}{\rho_2}\bigg)M_2- \Big(\frac{p^*}{\rho_2}+gh_2\Big)h_{2t}\,. 
 \end{array} 
\end{equation}
Then we transform the energy  equation of the system:
\[ \begin{array}{l}\displaystyle 
    \bigg(\frac{u_2^2}{2}+gH_0+\frac{p^*}{\rho_2}\bigg)\big(h_{2t}+(u_2h_2)_x\big)+
    u_2h_2\Big(u_{2t}+u_2u_{2x}+\frac{1}{\rho_2}p^*_x\Big) \\[3mm]\displaystyle
    \quad\quad =-\bigg(\frac{u^2}{2}\Big|_{z=z_2+0} +gH_0+\frac{p^*}{\rho_2}\bigg)M_2 
   \end{array} \]
Using the first and second equations in (\ref{eq:upper-layer}), we get
\begin{equation} \label{eq:compat-2}  
  \frac{M_2}{2}(u_2-u)^2\big|_{z=z_2+0}=0.
\end{equation}

Similarly, integrating equations (\ref{eq:Euler}), (\ref{eq:energy}) and using boundary conditions (\ref{eq:BC}), (\ref{eq:BC-interfaces}) and the pressure representation (\ref{eq:pressure-out}), we obtain the depth-averaged equations for the lower layer:
\begin{equation} \label{eq:lower-layer} 
 \begin{array}{l}\displaystyle 
  h_{1t}+(u_1 h_1)_x=-M_1, \\[3mm]\displaystyle 
  (u_1 h_1)_t+ \Big(u_1^2 h_1+\frac{gh_1^2}{2}+ (p^*+g\rho_2 h_2 +g\bar{\rho}\eta)\frac{h_1}{\rho_1} \Big)_x \\[3mm]\displaystyle 
  \quad =-M_1 u\big|_{z=z_1-0}+\frac{1}{\rho_1}(p^*+g\rho_2 h_2+ g\bar{\rho}\eta)h_{1x} -gh_1 Z_x\,, \\[3mm]\displaystyle
  \bigg(\frac{u_1^2 h_1}{2}+gZh_1+\frac{gh_1^2}{2}\bigg)_t+ \bigg(\frac{u_1^3 h_1}{2}+ \Big(\frac{p^*}{\rho_1}+gh_2\frac{\rho_2}{\rho_1}+ g\eta\frac{\bar{\rho}}{\rho_1}+gz_1\Big) u_1 h_1 \bigg)_x \\[3mm]\displaystyle 
   \quad = -\frac{M_1 u^2}{2}\Big|_{z=z_1-0} -gz_1 M_1- (M_1+h_{1t}) \bigg(\frac{p^*}{\rho_1}+ gh_2 \frac{\rho_2}{\rho_1}+ g\eta \frac{\bar{\rho}}{\rho_1}\bigg)+ gh_1 Z_t.
 \end{array}
\end{equation}
Taking into account the first two equations of (\ref{eq:lower-layer}),  the energy  equation yields:
\begin{equation} \label{eq:compat-1}  
  \frac{M_1}{2}(u_1-u)^2\big|_{z=z_1-0}=0.
\end{equation}

Suppose that the variables $M_1$ and $M_2$ are not identically equal to zero. Then, by virtue of (\ref{eq:compat-1}) and (\ref{eq:compat-2}), the compatibility condition between the energy, momentum and mass equations for flows in the outer layers gives us only one possibility
\begin{equation} \label{eq:compat}  
  u|_{z=z_1-0}=u_1, \quad u|_{z=z_2+0}=u_2. 
\end{equation}
Below we assume that conditions (\ref{eq:compat}) are satisfied.

\subsection{Depth-averaged equations for the intermediate non-homogeneous shear  layer}

As we already said  above, the velocity and density profiles at the internal interfaces are supposed to be  continuous, i.e.
\begin{equation} \label{eq:BC-u-rho}  
  (u,\rho)\big|_{z=z_1+0}=(u_1,\rho_1), \quad (u,\rho)\big|_{z=z_2-0}=(u_2,\rho_2). 
\end{equation}
Integrating the first three equations in system (\ref{eq:Euler}) and energy equation (\ref{eq:energy}) with respect to $z$ over the intermediate layer thickness and using boundary conditions (\ref{eq:BC-interfaces}), (\ref{eq:BC-rho}),  (\ref{eq:compat}), and (\ref{eq:BC-u-rho}), we get:
\begin{equation} \label{eq:ML-1}  
 \begin{array}{l}\displaystyle 
  \eta_t+(\bar{u}\eta)_x=M_1+M_2, \quad (\bar{\rho}\eta)_t+ \bigg(\int\limits_{z_1}^{z_2}u\rho\,dz\bigg)_x= \rho_1 M_1+ \rho_2 M_2, \\[3mm]\displaystyle
  \bigg(\int\limits_{z_1}^{z_2}u\rho\,dz\bigg)_t+ \bigg(\int\limits_{z_1}^{z_2}(u^2\rho+p)\,dz\bigg)_x \\[3mm]\displaystyle 
  \quad\quad\quad\quad\quad\quad\quad\quad =M_1 \rho_1 u_1+ M_2 \rho_2 u_2 +z_{2x}p\big|_{z=z_2}-z_{1x}p\big|_{z=z_1}\,, \\[3mm]\displaystyle
  \bigg(\int\limits_{z_1}^{z_2}E\,dz\bigg)_t+ \bigg(\int\limits_{z_1}^{z_2}(E+p)u\,dz\bigg)_x= M_2(E+p)\big|_{z=z_2} \\[3mm]\displaystyle 
  \quad\quad\quad\quad\quad\quad\quad\quad +M_1(E+p)\big|_{z=z_1}- z_{2t}p\big|_{z=z_2}+z_{1t}p\big|_{z=z_1}\,,
 \end{array}
\end{equation}
where $E=\rho(u^2/2+gz)$ is the long-wave energy. As before, we need to estimate the integrals equations (\ref{eq:ML-1}). 

Suppose that the stratification is weak and, following \cite{Tesh07}, we represent the dimensionless density in the form
\begin{equation} \label{eq:rho-repres}   
  \rho=\rho_c+\delta\tilde{\rho}(t,x,z), 
\end{equation} 
where $\rho_c=(\rho_1+\rho_2)/2$ and $\delta$ is a small parameter such that $\varepsilon^2\ll\delta\ll 1$. It is easy to see that, by virtue of the first two equations of system (\ref{eq:Euler}), the variable $\tilde{\rho}_z$ satisfies the equation 
\begin{equation} \label{eq:rho-der-z}   
 \frac{d\tilde{\rho}_z}{dt} = u_x \tilde{\rho}_z- u_z \tilde{\rho}_x.
\end{equation} 
Here $d/dt=\partial_t+u\partial_x+w\partial_z$ is the material derivative. 
Neglecting the terms $O(\varepsilon^2)$ in system (\ref{eq:Euler}) and taking into account representation (\ref{eq:rho-repres}), we can derive the following equation for the long-wave vorticity $u_z$:
\begin{equation} \label{eq:u-der-z}   
  \frac{du_z}{dt}= \frac{\delta p_x}{\rho^2}\tilde{\rho}_z+ \frac{\delta g}{\rho}\tilde{\rho}_x. 
\end{equation} 
Equations (\ref{eq:rho-der-z}) and (\ref{eq:u-der-z}) imply: if initially the variables $u_z$ and $\tilde{\rho}_z$ are small: $u_z=O(\gamma)$ and $\tilde{\rho}_z=O(\gamma)$ with $\varepsilon^2\ll\gamma\ll 1$, then for any time 
\begin{equation} \label{eq:der-est} 
  u_z=O(\gamma+\delta), \quad \tilde{\rho}_z=O(\gamma+\delta). 
\end{equation}  
The proof is a consequence of the linearity of equations for $u_z$ and $\tilde{\rho}_z$ (see \cite{Tesh07} for details). 

Formulae (\ref{eq:der-est}) yield the estimates
\begin{equation} \label{eq:u-rho-est}  
  |u-\bar{u}|=O(\gamma+\delta), \quad | \rho-\bar{\rho}|=O((\gamma+\delta)\delta).
\end{equation}  
Indeed, for any $z$ belonging to the interval $(z_1,z_2)$ we have
\[ \begin{array}{l}\displaystyle
   |u-\bar{u}|= \bigg| \int\limits_{z_1}^z u_z\,dz- \frac{1}{\eta} \int\limits_{z_1}^{z_2} 
   \bigg(\int\limits_{z_1}^{z'} u_z\,dz\bigg)\,dz'\bigg| \\[3mm]\displaystyle 
  \quad\quad\quad \leq \max|u_z| \bigg(\int\limits_{z_1}^z
   \,dz+ \frac{1}{\eta}\int\limits_{z_1}^{z_2} (z'-z_1)\,dz'\bigg) \leq \frac{3\eta}{2}\max|u_z|. 
  \end{array} \]
Taking into account representation (\ref{eq:rho-repres}), we similarly obtain the second estimate in (\ref{eq:u-rho-est}).

Using the  identities $\rho=\bar{\rho}+(\rho-\bar{\rho})$, $u=\bar{u}+(u-\bar{u})$ and formulae (\ref{eq:u-rho-est}), we obtain the following asymptotic estimates of integrals in (\ref{eq:ML-1}):
\[ \begin{array}{l}\displaystyle 
    \int\limits_{z_1}^{z_2} \rho u\,dz = \bar{u}\bar{\rho}\eta+  O((\gamma+\delta)^2\delta), \quad \int\limits_{z_1}^{z_2} \rho u^2\,dz= (\bar{u}^2+q^2) \bar{\rho}\eta+O((\gamma+\delta)^2\delta), \\[3mm]\displaystyle 
    \int\limits_{z_1}^{z_2} \rho u^3\,dz = (\bar{u}^2+3q^2)\bar{u}\bar{\rho}\eta+ O((\gamma+\delta)^2\delta), \quad \int\limits_{z_1}^{z_2} z\rho\,dz= \Big(Z+h_1+\frac{\eta}{2}\Big)\bar{\rho}\eta + O((\gamma+\delta)\delta),  
   \end{array} \]
   with the classical definition of $q^2$:
   \begin{equation}
   q^2=\frac{1}{\eta} \int\limits_{z_1}^{z_2}(u-\bar{u})^2 dz.
   \label{q_squared}
   \end{equation}
In view of (\ref{eq:pressure-gen}) and (\ref{eq:u-rho-est}) the pressure in this layer is  
\[ p=p^*+\rho_2 gh_2+g\bar{\rho}(z_2-z)+O((\gamma+\delta)\delta) \]
and, consequently, 
\[ \int\limits_{z_1}^{z_2} (p+g\rho z)u\,dz= (p^*+g\rho_2 h_2+g\, \bar{\rho}(Z+h_1+\eta)) 
   \bar{u}\eta +O((\gamma+\delta)\delta). \]
Note that these estimates are similar to those obtained in \cite{GLCh19}. Neglecting the terms of order $O((\gamma+\delta)\delta)$ in the previous integrals, we present system (\ref{eq:ML-1}) in the form:
\begin{equation} \label{eq:ML-2} 
 \begin{array}{l}\displaystyle 
  \eta_t+(\bar{u}\eta)_x=M_1+M_2, \quad (\bar{\rho}\eta)_t+(\bar{u}\bar{\rho}\eta)_x=M_1\rho_1+M_2\rho_2, \\[3mm]\displaystyle
  (\bar{u}\bar{\rho}\eta)_t+ \Big((\bar{u}^2+q^2)\bar{\rho}\eta+ (p^*+g\rho_2 h_2)\eta+ \frac{g\bar{\rho}\eta^2}{2}\Big)_x \\[3mm]\displaystyle 
  \quad\quad\quad =M_1\rho_1 u_1+ M_2\rho_2 u_2+ (p^*+g\rho_2 h_2)\eta_x- g\bar{\rho}\eta(Z+h_1)_x, \\[3mm]\displaystyle
  \bigg(\frac{(\bar{u}^2+q^2)}{2}\bar{\rho}\eta+ g(Z+h_1+\eta)\bar{\rho}\eta- \frac{g\bar{\rho}\eta^2}{2} \bigg)_t+ \bigg(\frac{(\bar{u}^2+3q^2)}{2}\bar{u}\bar{\rho}\eta \\[4mm]\displaystyle
  \quad +(p^*+g\rho_2 h_2) \bar{u}\eta+g(Z+h_1+\eta)\bar{u}\bar{\rho}\eta\bigg)_x= M_1\rho_1\Big(\frac{u_1^2}{2}+g(Z+h_1)\Big) \\[3mm]\displaystyle  
  \quad\quad +M_2\rho_2\Big(\frac{u_2^2}{2}+g(Z+h_1+\eta)\Big)+ (p^*+g\rho_2 h_2)(M_1+M_2-\eta_t)
  \\[3mm]\displaystyle  
  \quad\quad\quad +g(M_1+Z_t+h_{1t})\bar{\rho}\eta- d.
 \end{array} 
\end{equation}
To account for the energy dissipation in the last equation of (\ref{eq:ML-2}) an extra term $d$ has been added (formula for $d$ will be proposed later).

\subsection{Differential consequence of the energy equation}

Let us derive an equation for the variable $q$ measuring the distortion of the velocity profile. For this, we first note that the first three equations (\ref{eq:ML-2}) imply:
\begin{equation} \label{eq:conseq-rho-u}
 \begin{array}{l}\displaystyle
  \bar{\rho}_t+\bar{u}\bar{\rho}_x=\frac{1}{\eta}\Big((\rho_1-\bar{\rho})M_1+ (\rho_2-\bar{\rho})M_2\Big), \\[3mm]\displaystyle
  \bar{u}_t+ \bar{u}\bar{u}_x+2qq_x+ g h_{1x}+ \Big(g+\frac{q^2}{\eta}\Big)\eta_x+ \frac{g\rho_2}{\bar{\rho}}h_{2x}+ \frac{1}{\bar{\rho}}\Big(q^2+\frac{g\eta}{2}\Big)\bar{\rho}_x +\frac{1}{\bar{\rho}}p^*_x \\[3mm]\displaystyle
  \quad\quad\quad =-gZ_x+ \frac{1}{\bar{\rho}\eta}\Big((u_1-\bar{u})\rho_1 M_1+ (u_2-\bar{u})\rho_2 M_2\Big). 
 \end{array}
\end{equation} 
The non-conservative form of the last equation in (\ref{eq:ML-2}) is
\[ \begin{array}{l}\displaystyle 
    \Big(\frac{\bar{u}^2+q^2}{2}+g(Z+h_1+\eta)\Big) \big((\bar{\rho}\eta)_t+ 
    (\bar{u}\bar{\rho}\eta)_x\big)+ q\bar{\rho}\eta \big(q_t+(\bar{u}q)_x\big) \\[3mm]\displaystyle
    \quad +\bar{u}\bar{\rho}\eta \Big(\bar{u}_t+ \bar{u}\bar{u}_x+2qq_x
    +g h_{1x}+ \Big(g+\frac{q^2}{\eta}\Big)\eta_x+ \frac{g\rho_2}{\bar{\rho}}h_{2x}+ \frac{1}{\bar{\rho}}\Big(q^2+\frac{g\eta}{2}\Big)\bar{\rho}_x +\frac{1}{\bar{\rho}}p^*_x \Big) \\[3mm]\displaystyle  
    \quad\quad -\frac{g\eta^2}{2}\big(\bar{\rho}_t+\bar{u}\bar{\rho}_x\big)+ g\bar{u}\bar{\rho}\eta Z_x= M_1\rho_1\Big(\frac{u_1^2}{2}+g(Z+h_1)\Big) \\[3mm]\displaystyle
    \quad\quad\quad +M_2\rho_2\Big(\frac{u_2^2}{2}+g(Z+h_1+\eta)\Big)+ M_1 g\bar{\rho}\eta- d. 
   \end{array} \]
Taking into account relations (\ref{eq:conseq-rho-u}) and the second equation in (\ref{eq:ML-2}), we obtain
\begin{equation} \label{eq:conseq-q} 
 \begin{array}{l}\displaystyle 
  q\bar{\rho}\eta\big(q_t+(\bar{u}q)_x\big)= \frac{M_1 \rho_1}{2} \big((u_1-\bar{u})^2-q^2-g\eta\big) \\[3mm]\displaystyle \quad\quad\quad +\frac{M_2 \rho_2}{2} \big((u_2-\bar{u})^2-q^2+g\eta\big)+ \frac{g\bar{\rho}\eta}{2}(M_1-M_2)-d. 
 \end{array} 
\end{equation} 
From this consequence and the mass equation for $\eta$ one can derive the transport equation for $q/\eta$ which can be interpreted as the evolution equation for the flow vorticity in the intermediate layer. A priori, the vorticity can change its sign during the flow evolution.

\subsection{Final three-layer system}

The final system can be written as 
\begin{equation} \label{eq:CL-3l}  
 \begin{array}{l}\displaystyle
  h_{1t}+(u_1 h_1)_x=-M_1, \quad \eta_t+(\bar{u} \eta)_x= M_1+M_2, \quad h_{2t}+(u_2 h_2)_x=-M_2, \\[3mm]\displaystyle
  u_{1t}+ \bigg(\frac{u_1^2}{2}+ gh_1+ g\eta\frac{\bar{\rho}}{\rho_1}+ gh_2\frac{\rho_2}{\rho_1}+ \frac{p^*}{\rho_1}\bigg)_x= -gZ_x, \\[4mm]\displaystyle
  u_{2t}+\bigg(\frac{u_2^2}{2}+ \frac{p^*}{\rho_2}\bigg)_x=0, \quad 
  Q_t+ \bigg(\rho_1 u_1^2 h_1+ (\bar{u}^2+q^2)\bar{\rho}\eta+ \rho_2 u_2^2 h_2+ Hp^* \\[3mm]\displaystyle 
  \quad +\frac{gh_1^2\rho_1}{2}+ \frac{g\eta^2\bar{\rho}}{2}+ \frac{gh_2^2\rho_2}{2}
  +g\bar{\rho}h_1\eta + g\rho_2h_2(h_1+\eta)\bigg)_x \\[3mm]\displaystyle
  \quad\quad =-(gh_1\rho_1+g\eta\bar{\rho}+gh_2\rho_2+p^*)Z_x, 
  (\bar{\rho}\eta)_t+(\bar{u}\bar{\rho}\eta)_x=M_1\rho_1+M_2\rho_2, \\[3mm]\displaystyle
  q_t+(\bar{u}q)_x= \frac{M_1 \rho_1}{2q\eta\bar{\rho}} \big((u_1-\bar{u})^2-q^2-g\eta\big) \\[3mm]\displaystyle 
  \quad\quad +\frac{M_2 \rho_2}{2q\eta\bar{\rho}} \big((u_2-\bar{u})^2-q^2+g\eta\big) +\frac{g}{2q}(M_1-M_2)-\frac{d}{q\bar{\rho}\eta},
\end{array}
\end{equation} 
where $Q=\rho_1 u_1 h_1+ \bar{\rho}\bar{u}\eta+ \rho_2 u_2 h_2$ is the total mass discharge,  and $H=H_0-Z$ is the total fluid thickness. To represent the depth-averaged equations in the form of balance laws, we combined the momentum equations in the outer layers and intermediate layer into the total momentum equation for $Q$. 

Instead of the cumbersome total energy equation, we use its differential consequence for the $q$ variable. This replacement of the balance law does not affect the construction of solutions in the class of continuous flows, but in the event of discontinuities, the solutions may differ. However, in the case of small jump amplitude this difference is negligible. An example of such an approach is  shown by \cite{LipLiapCh21} for the case of  compressible flows. This  difference  manifests itself in a fairly small region, outside of which the solutions are close or even almost coincide. The appropriateness  of using the equation for the velocity distortion $q$ is also underlined by \cite{GLCh16, GLCh19} where free surface two-layer flows taking into account mixing were considered.

To close the model it is necessary to determine the mass entrainment  terms  $M_1$, $M_2$,  and the energy dissipation term $d$. Since stratification is considered weak, we suppose that the entrainment of fluid from the outer layers into the intermediate one is symmetric. Following \cite{GLCh16, GLCh19, LiapDG18}, we take the entrainment velocities $M_1$ and $M_2$ in the form
\begin{equation} \label{eq:M1_M2} 
M_1=M_2=\sigma q, \quad \sigma={\rm const}>0. 
\end{equation} 
The interpretation of $\sigma$ parameter comes  from the theory of plane  turbulent shear. More, exactly,  $2\sigma $ is the ratio of the shear stress to the  turbulent energy $k_T$,  $k_T\approx q^2/2$. This ratio is approximately constant and equal to $0.3$ (see \cite{Pope_2000}, Table 5.4, p.\,157). Thus, in the following, we always take $\sigma=0.15$. A detailed justification of the entrainment terms can  be found in~\cite{GLCh16}. 

To account for the energy transfer  from large scale eddies to small scale eddies  we added the extra term $d$ in the last equation of (\ref{eq:ML-2}): 
\begin{equation} \label{eq:dissip-term}  
  d=\frac{\sigma\kappa}{2}\bar{\rho}|q|^3, \quad \kappa={\rm const}>0.  
\end{equation}
Such a  dissipation term is classical in the theory of turbulent flows of homogeneous fluids. Indeed, the rate of turbulent energy dissipation  is usually  written as  (see \cite{Pope_2000},  p.\,244): 
\begin{equation}
\epsilon=\frac{k_T^{3/2}}{L_{11}}\left(\frac{L_{11}}{L}\right), 
\end{equation}
where $L_{11}$ is the length scale of the energy-containing eddies, and $\varkappa=L_{11}/L$ is a dimensionless parameter which tends asymptotically for high Reynolds numbers to a constant value (see \cite{Pope_2000}, p. 244--245). In our  case,  the horizontal velocity  in the intermediate shear layer  can be approximated  by a linear profile :  $u=\bar{u}+\omega \left(z-(z_1+z_2)/2\right)$. Here $\omega ={\rm const}$ is the component of the horizontal vorticity.  Since the velocity profile is symmetric with respect to $\bar{u}$, one  can conclude  that the size of the energy-containing eddies is $\eta/2$. Hence,   
\begin{equation}
\epsilon=\frac{k_T^{3/2}}{L_{11}}\left(\frac{L_{11}}{L}\right)=\frac{\vert q\vert^3}{2\sqrt{2}L_{11}}\left(\frac{L_{11}}{L}\right)=\frac{\vert q\vert^3}{\eta\sqrt{2}}\left(\frac{L_{11}}{L}\right)= \frac{\vert q\vert^3}{2\eta}\left(\frac{\sqrt{2}L_{11}}{L}\right)
\end{equation}
Thus
\begin{equation}
\frac{L_{11}}{L}=\frac{\kappa \sigma}{\sqrt{2}}.
\end{equation}
The experimentally observed  data for the  ratio $L_{11}/L$ is  (\cite{Pope_2000},  p.\,245): 
\begin{equation}
0.8>\frac{L_{11}}{L}=\frac{\kappa \sigma}{\sqrt{2}}>0.43. 
\end{equation}
For $\sigma\approx 0.15$ it implies the following estimation for $\kappa$ :
\begin{equation}
\simeq 8>\kappa>\simeq 4. 
\end{equation}
Finally, $\kappa$ is the  only phenomenological parameter in our model. As we have observed in our numerical experiments,  the results do not depend too much on the specific choice of $\kappa \in (4,8)$. 

Thus, equations (\ref{eq:CL-3l}) with the additional relation $h_1+\eta+h_2=H_0-Z$ form a closed system for nine unknown functions $h_1$, $\eta$, $h_2$, $u_1$, $\bar{u}$, $u_2$, $\bar{\rho}$, $q$ and $p^*$. 

\section{Governing equations in the Boussinesq approximation}
\label{sec:Bouss_model}

Let us return back to dimensional variables. The depth-averaged equations will not change if we drop the `hats' on dimensional variables. Further we apply the Boussinesq approximation assuming that the ratio $(\rho_1-\rho_2)/\rho_2\ll 1$ is negligible, but not the  buoyancy terms 
\[ b=g\frac{\rho_1-\rho_2}{\rho_2}, \quad \bar{b}=g\frac{\bar{\rho}-\rho_2}{\rho_2}\,. \]
Since the fourth, sixth, and seventh equations in (\ref{eq:CL-3l}) admit the representation
\[ \begin{array}{l}\displaystyle
    u_{1t}+ \bigg(\frac{u_1^2}{2}+ \frac{p^*}{\rho_1}+ gh_1\frac{\rho_1-\rho_2}{\rho_1}+ g\eta\frac{\bar{\rho}-\rho_2}{\rho_1}\bigg)_x= -gZ_x\frac{\rho_1-\rho_2}{\rho_1}, \\[3mm]\displaystyle
    Q_t+ \bigg(\rho_1 u_1^2 h_1+ (\bar{u}^2+q^2)\bar{\rho}\eta+ \rho_2 u_2^2 h_2+ 
    \frac{g h_1^2}{2}(\rho_1-\rho_2)+ gh_1\eta(\bar{\rho}-\rho_2) \\[3mm]\displaystyle
    \quad\quad\quad +\frac{g\eta^2}{2}(\bar{\rho}-\rho_2) + Hp^* \bigg)_x= 
    -(gh_1(\rho_1-\rho_2)+ g\eta(\bar{\rho}-\rho_2)+p^*)Z_x, \\[3mm]\displaystyle
    \big((\bar{\rho}-\rho_2)\eta\big)_t+ \big((\bar{\rho}-\rho_2)\bar{u}\eta\big)_x = (\rho_1-\rho_2)M_1
   \end{array} \]
and the ratios $1/\rho_1$ and $1/\bar{\rho}$ can be approximated as
\[ \frac{1}{\rho_1}= \frac{1}{\rho_2+(\rho_1-\rho_2)} \approx \frac{1}{\rho_2}-
   \frac{\rho_1-\rho_2}{\rho_2^2}\,, \quad \frac{1}{\rho_1}\approx \frac{1}{\rho_2}-
   \frac{\bar{\rho}-\rho_2}{\rho_2^2}, \]
the governing equations take the form
\begin{equation} \label{eq:CL-3l-Boussinesq}  
 \begin{array}{l}\displaystyle
   h_{1t}+(u_1 h_1)_x=-\sigma q, \quad \eta_t+(\bar{u} \eta)_x= 2\sigma q, \quad 
   h_{2t}+(u_2 h_2)_x=-\sigma q, \\[3mm]\displaystyle
   u_{1t}+ \bigg(\frac{u_1^2}{2}+ \frac{p^*}{\rho_2}+ bh_1+ \bar{b}\eta \bigg)_x= -bZ_x, \quad 
   u_{2t}+\bigg(\frac{u_2^2}{2}+\frac{p^*}{\rho_2}\bigg)_x=0, \\[4mm]\displaystyle
   \bar{Q}_t+ \bigg(u_1^2 h_1+ (\bar{u}^2+q^2)\eta+ u_2^2 h_2+ \frac{bh_1^2}{2}+ 
   \bar{b}h_1\eta+ \frac{\bar{b}\eta^2}{2}+ \frac{Hp^*}{\rho_2}\bigg)_x \\[4mm]\displaystyle
   \quad\quad =-\bigg(bh_1+\bar{b}\eta+ \frac{p^*}{\rho_2} \bigg)Z_x, \quad
   (\bar{b}\eta)_t+ (\bar{u}\bar{b}\eta)_x=\sigma q b, \quad q_t+(\bar{u}q)_x=\varphi. 
 \end{array}
\end{equation} 
Here we have already taken into account formulae (\ref{eq:M1_M2}), (\ref{eq:dissip-term}) and used the notation
\[ \varphi=\frac{\sigma}{2\eta}\Big((u_1-\bar{u})^2+(u_2-\bar{u})^2- 
   (2+\kappa\,{\rm sign}\, q)q^2-b\eta\Big) \]
and $\bar{Q}=u_1 h_1+\bar{u}\eta+u_2 h_2$. In view of the first three equations  (\ref{eq:CL-3l-Boussinesq}) we have $\bar{Q}_x=-H_t=Z_t$. Therefore, the variable $\bar{Q}$ is known (up to an arbitrary function of time).

One can derive from (\ref{eq:ML-2}), (\ref{eq:lower-layer}) and (\ref{eq:upper-layer})   the conservation equation of the total energy of a three-layer flow in the  Boussinesq approximation:
\begin{equation} \label{eq:energy-bouss}  
	\begin{array}{l}\displaystyle 
	\bigg(\frac{u_1^2 h_1}{2}+ \frac{(\bar{u}^2+q^2)\eta}{2}+ \frac{u_2^2 h_2}{2}+ \frac{bh_1^2}{2}+ \bar{b}\eta\Big(h_1+\frac{\eta}{2}\Big)\bigg)_t+ \bigg(\frac{u_1^3 h_1}{2}  \\[3mm]\displaystyle
	\quad +\frac{(\bar{u}^2+3q^2)\bar{u}\eta}{2}+\frac{u_2^3 h_2}{2}  +(bh_1+\bar{b}\eta)u_1 h_1 +\bar{u}\bar{b}\eta(h_1+\eta) +p^*\bar{Q}\bigg)_x \\[3mm]\displaystyle 
	\quad\quad = p^* Z_t- (u_1 h_1 b +v\eta\bar{b})Z_x- \frac{\sigma\kappa}{2}|q|^3\,.
	\end{array}
\end{equation} 
As mentioned above, we have replaced this cumbersome equation by its more convenient differential consequence for the variable $q$. {\it A priori}, this procedure  is equivalent for smooth solutions, but not for discontinuous solutions.  In the next section, we will show that such a change in the conservation law does not significantly affect the structure of  discontinuous solutions.

It is convenient to derive the following consequences of system (\ref{eq:CL-3l-Boussinesq}) for variables $\bar{b}$, $s=q/\eta$ and $\bar{u}$:
\begin{equation} \label{eq:buq} 
 \begin{array}{l}\displaystyle
  \bar{b}_t+\bar{u}\bar{b}_x=\frac{\sigma q}{\eta}(b-2\bar{b}), \quad
  s_t+\bar{u}s_x= \frac{1}{\eta}\Big(\varphi-\frac{2\sigma q^2}{\eta}\Big), \\[3mm]\displaystyle  
  \bar{u}_t+ \bar{u}\bar{u}_x+ 2qq_x+ \bar{b}h_{1x}+ \Big(\bar{b}+\frac{q^2}{\eta}\Big)\eta_x+ 
  \frac{\eta}{2}\bar{b}_x+ \frac{1}{\rho_2}p^*_x \\[3mm]\displaystyle
  \quad\quad\quad =\frac{\sigma q}{\eta}(u_1+u_2-2\bar{u})- \bar{b}Z_x.
 \end{array}
\end{equation} 
Note that these equations can be obtained from  (\ref{eq:conseq-rho-u}) and (\ref{eq:conseq-q}) by passing to the Boussinesq approximation and using formulae (\ref{eq:M1_M2}) and (\ref{eq:dissip-term}). We write now the  first equation of (\ref{eq:buq}):  
\begin{equation}
\bar{b}_t+\bar{u}\bar{b}_x=\frac{\sigma q}{\eta}(b-2\bar{b}). 
\label{eq:buq_1} 
\end{equation}
The buoyancy $\bar b$ has  the following  property: if,  initially,   $\bar b >b/2$, then this property is valid any time. Moreover,  $\bar b $ is always positive. These properties follow directly from (\ref{eq:buq_1}). In particular, this implies that for any time  $\rho_2<\bar \rho<\rho_1$, i.e. the stratification stays always stable. 

Let us eliminate the pressure $p^*$ on the upper lid. For this, we introduce new variables 
\[ r=u_1-u_2, \quad R=\bar{Q}-u_2 H=(u_1-u_2)h_1+(\bar{u}-u_2)\eta \]
and note that the variable $u_2H$ satisfies equation
\[ (u_2 H)_t+\bigg(\frac{u_2^2H}{2}+\frac{Hp^*}{\rho_2}\bigg)_x= 
   -\Big(Z_t+\frac{u_2}{2}Z_x\Big)u_2- \frac{p^*}{\rho_2}Z_x. \]
Subtracting this balance law from the sixth equation (\ref{eq:CL-3l-Boussinesq}), as well as the fifth from the fourth, we get an evolutionary system.

Thus, to define six unknowns $(h_1,h_2,r,R,\bar{b},q)$ we obtain the closed system of balance laws
\begin{equation} \label{eq:model}  
 \begin{array}{l}\displaystyle
  h_{1t}+(u_1 h_1)_x=-\sigma q, \quad \eta_t+(\bar{u}\eta)_x=2\sigma q, \\[3mm]\displaystyle
  (\bar{b}\eta\big)_t+ (\bar{u}\bar{b}\eta)_x=\sigma q b, \quad
  r_t+ \bigg(\frac{u_1^2-u_2^2}{2}+ bh_1+ \bar{b}\eta \bigg)_x= -bZ_x, \\[3mm]\displaystyle
  R_t+ \bigg(u_1^2 h_1+ (\bar{u}^2+q^2)\eta+ \Big(h_2-\frac{H}{2}\Big)u_2^2+ 
  \frac{bh_1^2}{2}+ \bar{b}h_1\eta+ \frac{\bar{b}\eta^2}{2}\bigg)_x \\[3mm]\displaystyle
  \quad\quad\quad =u_2 Z_t +\Big(\frac{u_2^2}{2}- bh_1-\bar{b}\eta\Big)Z_x, \quad
  q_t+(\bar{u}q)_x=\varphi,
 \end{array}
\end{equation} 
where the variables $h_2$, $u_1$, $u_2$ and $\bar{u}$ can be expressed as
\[ h_2=H-h_1-\eta, \quad u_2=\frac{\bar{Q}-R}{H}, \quad 
   u_1=r+u_2, \quad \bar{u}=\frac{R-rh_1}{\eta}+u_2. \]
System (\ref{eq:model}) describes non-stationary three-layer hydrostatic flows with mixing in the Boussinesq approximation. 

\subsection{Characteristics of equations (\ref{eq:model})}
Let us rewrite system (\ref{eq:model}) in the form 
\begin{equation} \label{eq:model-vec}  
  \bU_t+\bA\bU_x=\bF, 
\end{equation} 
where $\bU=(\bar{b}, s, h_1, \eta, r, R)^{\rm T}$ is the vector of unknowns, $\bF$ is the right-hand  which doesn't contain derivatives of $\bU$, and 
\[ \bA= \begin{pmatrix} 
    \bar{u} & 0       & 0   & 0       & 0    & 0      \\ 
    0       & \bar{u} & 0   & 0       & 0    & 0      \\
    0       & 0       & u_1 & 0       & h_1  & -h_1/H \\
    0       & 0       & -r  & u_2     & -h_1 & 1-\eta/H \\
    \eta    & 0       & b   & \bar{b} & u_1  & -r/H \\
    a_1     & a_2     & a_3 & a_4     & a_5  & a_6 \\
\end{pmatrix} \]
is the $6\times 6$ matrix.  The last row of the matrix is
\[ \begin{array}{l}\displaystyle 
   a_1=\Big(h_1+\frac{\eta}{2}\Big)\eta, \quad a_2=2\eta^2 q, \quad 
   a_3=(u_1-\bar{u})^2-(u_2-\bar{u})^2+bh_1+\bar{b}\eta, \\[2mm]\displaystyle 
   a_4=3q^2-(u_2-\bar{u})^2+(h_1+\eta)\bar{b}, \quad a_5=2(u_1-\bar{u})h_1, \quad
   a_6=u_2+2\Big(\bar{u}-\frac{\bar{Q}}{H}\Big).
   \end{array} \]

The eigenvalues of $\bA(\bU)$ are the characteristic velocities of system (\ref{eq:model-vec}). The sixth order polynomial equation ${\rm det}(\bA-\lambda\bI)=0$ has the root $\lambda=\bar{u}$ of multiplicity two. To determine the remaining four roots, one has to solve the equation
\begin{equation} 
\label{eq:chi}  
 \begin{array}{l}\displaystyle 
   \chi(\lambda)=\big((u_1-\lambda)^2-(b-\bar{b})h_1\big) \big((u_2-\lambda)^2-\bar{b}h_2\big)\eta \\[2mm]\displaystyle
  \quad +\big((\bar{u}-\lambda)^2-3q^2\big) \Big(\big((u_1-\lambda)^2-(b-\bar{b})h_1\big)h_2+ \big((u_2-\lambda)^2-\bar{b}h_2\big)h_1\Big)=0.
 \end{array} 
\end{equation}

Consider the characteristic polynomial $\chi(\lambda)$ in the case of equal velocities $u_1=u_2=\overline{u}$. Due to the  Galilean invariance of the governing equations in the case  $\mathbf{F}=\mathbf{0}$,  one can always take  the depth averaged velocities vanishing. However, $q$ does not vanish. Then equation (\ref{eq:chi}) multiplied by $(h_1 \eta h_2)^{-1}$ is 
\begin{equation}\label{eq:polynomial_simpified} 
 \begin{array}{l}\displaystyle 
  \bigg(\frac{1}{h_1 h_2}+\frac{1}{ h_1 \eta}+\frac{1}{\eta h_2}\bigg)\lambda^4 -
  \bigg(\frac{\bar{b}}{h_1}+\frac{b-\bar{b}}{ h_2}+\frac{b}{\eta} +\frac{3q^2}{\eta} \Big(\frac{1}{h_1}+\frac{1}{h_2}\Big)\bigg)\lambda^2 \\[4mm]\displaystyle 
  \quad\quad\quad\quad\quad +\frac{3q^2 b}{\eta}+ \bar{b}(b-\bar{b})=0. 
 \end{array}
\end{equation} 
Let $\mu_1=\lambda_1^2$ and $\mu_2=\lambda_2^2$ be the roots of the corresponding bi-quadratic equation for $\mu=\lambda^2$. Since the stratification is stable, one has $b>\overline{b}>0$. Then one obviously has 
\[ \mu_1+\mu_2>0,\quad \mu_1 \mu_2>0. \]  

Let us first show that for $q=0$ all roots $\mu_i$ are positive. Indeed, for this it is sufficient  to show that the corresponding discriminant $\Delta$ is positive :  
\[ \begin{array}{l}\displaystyle 
    \Delta =\left(\frac{\bar{b}^2}{h_1^2}+ \frac{(b-\bar{b})^2}{h_2^2}+ \frac{b^2}{\eta^2}+\frac{2{\bar{b}}(b-\bar{b})}{h_1h_2}+
	\frac{2b(b-\bar{b})}{\eta h_2}+ \frac{2b\bar{b}}{h_1\eta} \right) \\[4mm]\displaystyle 
	\quad\quad\quad -4\bar{b}(b-\bar{b}) \left(\frac{1}{h_1 h_2}+\frac{1}{ h_1 \eta}+\frac{1}{\eta h_2}\right). 
   \end{array} \]
Let $\bz=(1/h_1,1/h_2, 1/\eta)^{\rm T}$. Then one can write
\[ \Delta =\bz^{\rm T}\cdot \bB \cdot \bz, \quad {\bB}=
	\left(\begin{array}{ccc}
		{\bar{b}}^2& -{\bar{b}}(b-{\bar{b}})& -{\bar{b}}(b-2{\bar{b}})\\
		-{\bar{b}}(b-{\bar{b}})& (b-{\bar{b}})^2& (b-{\bar{b}})(b-2{\bar{b})}\\
		-{\bar{b}}(b-2{\bar{b}})& (b-{\bar{b}})(b-2{\bar{b})}& b^2 \\
	\end{array}\right). \]
Both the second and third principal minors  of $\bB$ are identically zero, so the matrix is only non-negative definite: $\bB\ge 0$. The case $q\ne 0$  improves the situation: the corresponding matrix is always degenerate, but  first and second principal minor are positive.  Overall, the roots $\mu_i$ are positive, but they can coincide. This means that in the absence of a velocity shear, all eigenvalues are real. 

When the governing equations are hyperbolic  the concepts of supercritical and subcritical flow can be introduced.  We say that the flow is {\it  supercritical} if all roots $\lambda=\lambda_i$ of the characteristic equation (\ref{eq:chi}) are positive, and {\it subcritical} if there is at least one negative root.

\section{Stationary solutions}
\label{sec:stationary}

Stationary solutions of system (\ref{eq:CL-3l-Boussinesq}) (or (\ref{eq:model})) are determined by the equations
\begin{equation} \label{eq:st-3l}  
 \begin{array}{l}\displaystyle
  (u_1 h_1)'=-\sigma q, \quad (\bar{u}\eta)'=2\sigma q, \quad 
  (\bar{u}\eta\bar{b})'=\sigma q b, \\[3mm]\displaystyle
  (\bar{u}q)'=\varphi, \quad \Big(\frac{u_1^2-u_2^2}{2} +b h_1+ \bar{b}\eta\Big)'=-bZ', \\[3mm]\displaystyle
  \bar{u}\bar{u}'+2qq'+\bar{b}h_1'+ \Big(\bar{b}+\frac{q^2}{\eta}\Big)\eta'+ \frac{\eta}{2}\bar{b}'-u_2 u_2'= \frac{\sigma q}{\eta}(u_1-2\bar{u}+u_2)-\bar{b}Z'.
 \end{array}
\end{equation}
Here `prime' means the derivative with respect to $x$. Since $u_2=(\bar{Q}-u_1 h_1-\bar{u}\eta)/h_2$ and $h_2=H_0-Z-h_1-\eta$, equations (\ref{eq:st-3l}) form a closed system for unknowns $h_1$, $\eta$, $u_1$, $\bar{u}$, $q$ and $\bar{b}$. Let us rewrite this system in normal form
\begin{equation} \label{eq:st-3l-norm} 
 \begin{array}{l}\displaystyle 
  u_1'=-\frac{\sigma q+u_1 h_1'}{h_1}, \quad \bar{u}'=\frac{2\sigma q-\bar{u}\eta'}{\eta}, \quad 
  \quad \bar{b}'=\frac{\sigma q}{\bar{u}\eta} (b-2\bar{b}), \\[4mm]\displaystyle
  q'=\frac{\varphi-q\bar{u}'}{\bar{u}}, \quad \eta'=\frac{C_1-A_1 h_1'}{A_2}, \quad h_1'=\frac{B_2 C_1-A_2 C_2}{A_1 B_2-A_2^2}\,,
 \end{array} 
\end{equation}
where
\[ \begin{array}{l}\displaystyle 
    A_1=\frac{u_1^2}{h_1}+\frac{u_2^2}{h_2}-b, \quad A_2=\frac{u_2^2}{h_2}-\bar{b}, \quad 
    C_1=\eta\bar{b}' -\sigma q\Big(\frac{u_1}{h_1}-\frac{u_2}{h_2}\Big) -\Big(\frac{u_2^2}{h_2}-b\Big)Z',  \\[3mm]\displaystyle
    B_2=\frac{\bar{u}^2-3q^2}{\eta}+A_2, \quad C_2=\frac{\eta\bar{b}'}{2} -\frac{\sigma q}{\eta}\Big(u_1-4\bar{u}+\frac{h_2-\eta}{h_2}u_2\Big) +\frac{2q}{\bar{u}}\Big(\varphi-\frac{2\sigma q^2}{\eta}\Big)- A_2 Z'.
  \end{array} \]

We note that 
\[ \chi(0)=(A_1 B_2-A_2^2)h_1\eta h_2 \] 
and, consequently, if the denominator of the right-hand side in the last equation~(\ref{eq:st-3l-norm}) vanishes, then $\lambda=0$ is the root of characteristic equation~(\ref{eq:chi}). We say that a stationary flow is supercritical if $\chi(0)>0$, and subcritical if $\chi(0)<0$. 

To solve ODEs (\ref{eq:st-3l-norm}) numerically, we use the standard ode45 procedure of the MATLAB package that implements the fourth-order Runge--Kutta method. 

\subsection{Mixing layer formation} 

To construct a stationary solution to the mixing layer problem, it is necessary to determine the values of $\bar{u}$, $q$ and $\bar{b}$ as $\eta\to 0$. Without loss of generality, we assume that $\eta=0$ at $x=0$, and the values of the functions at this point are marked with the subscript `zero'. Velocities $u_{10}>0$, $u_{20}>0$ ($u_{10}\neq u_{20}$) and thicknesses $h_{10}>0$, $h_{20}>0$ for the outer layers are assumed to be specified. We also assume that there are finite limits of all functions and their derivatives. By virtue of the second, third, fourth, and sixth equations of system~(\ref{eq:st-3l}) for $\eta\to 0$, we obtain
\[ \begin{array}{l}\displaystyle 
  \eta'\to \frac{2\sigma q_0}{\bar{u}_0}, \quad \bar{b}_0=\frac{b}{2}, \quad (u_{10}-\bar{u}_0)^2+(u_{20}-\bar{u}_0)^2=(2+\kappa)q_0^2, \\[3mm]\displaystyle
  \eta'\to \frac{\sigma}{q_0}(u_{10}+u_{20}-2\bar{u}_0).
 \end{array} \]
This means that 
\begin{equation} \label{eq:q0} 
 q_0=\sqrt{(u_{10}+u_{20}-2\bar{u}_0)\bar{u}_0/2} 
\end{equation} 
($q_0>0$ since during the mixing layer formation, it expands) and $\bar{u}_0$ is the root of the quadric equation
\begin{equation} \label{eq:u0}  
  \bar{u}_0^2-\frac{(6+\kappa)(u_{10}+u_{20})}{2(4+\kappa)}\bar{u}_0+ 
  \frac{u_{10}^2+u_{20}^2}{4+\kappa}=0. 
\end{equation}
Variation of the dissipation parameter $\kappa$ allows us to obtain the unique root of this equation $\bar{u}_0>(u_{10}+u_{20})/2$ lying between the values of $u_{10}$ and $u_{20}$. This can always be achieved for not too large absolute value of the  relative velocity $|u_{20}-u_{10}|$.

These restrictions on the variables $\bar{u}_0$ and $q_0$ apply only to the initial stage of mixing, where the thickness of the interlayer is close to zero. If non-stationary equations (\ref{eq:model}) are used to calculate the evolution of the mixing layer, then instead of conditions (\ref{eq:q0}), (\ref{eq:u0}) we can set, for example $\bar{u}_0=(u_1+u_2)/2$, $q_0=0$. It will be shown below that when the solution of equations (\ref{eq:model}) with the indicated conditions for $\bar{u}_0$ and $q_0$ reaches the stationary regime, it coincides with the corresponding solution of stationary equations (\ref{eq:st-3l-norm}) everywhere, except for a small neighbourhood of the cross-section $x=0$. To construct a numerical solution of stationary or non-stationary equations, it is necessary to set a small but positive value of the interlayer thickness $\eta$ at $x=0$.

\subsection{The structure of stationary solutions}

The conservative form of equations (\ref{eq:model}) allows one to construct solutions with a hydraulic jump  that transforms a supercritical flow into a subcritical one. Across the jump the Rankine--Hugoniot relations are satisfied :
\begin{equation} \label{eq:R-Hugoniot-1}  
[Q_i]=0, \quad [Q_m]=0,  \quad [\bar{u}\eta\bar{b}]=0, \quad [J_s]=0, \quad [J]=0, \quad 
[W]=0, 
\end{equation}
where 
\begin{equation} \label{eq:R-Hugoniot-2} 
\begin{array}{l} \displaystyle 
Q_i=u_i h_i, \quad Q_m=\bar{u}\eta, \quad J_s=\bar{u}q, \quad J=\frac{u_1^2-u_2^2}{2}+bh_1+\bar{b}\eta, \\[3mm]\displaystyle
W=u_1^2 h_1+(\bar{u}^2+q^2)\eta+u_2^2\Big(h_2-\frac{H}{2}\Big)+ \frac{bh_1^2}{2}+ \Big(h_1+\frac{\eta}{2}\Big)\eta\bar{b} 
\end{array} 
\end{equation}
$i=1,2$ and the square brackets mean the jump of the corresponding quantities. These relations follow from conservation laws (\ref{eq:model}). By virtue of (\ref{eq:R-Hugoniot-1}),  variables defined by formulas (\ref{eq:R-Hugoniot-2}), as well as the buoyancy $\bar{b}$, are continuous through the jump. 

Let the supercritical flow ahead of  the jump at $x=x_s-0$ be known. Then behind the jump (values of the functions at $x=x_s\pm0$ are denoted with superscript `$\pm$') we have
\begin{equation} \label{eq:vel-plus}  
u_1^+=\frac{Q_1}{h_1^+}, \quad \bar{u}^+=\frac{Q_m}{\eta^+}, \quad u_2^+=\frac{Q_2}{h_2^+}, \quad q^+=\frac{J_s}{Q_m}\eta^+ \quad (h_2^+=H-h_1^+-\eta^+) 
\end{equation}
and the following system of algebraic equations for finding $h_1^+$, $\eta^+$:
\begin{equation} \label{eq:R-Hugoniot-3} 
\begin{array}{l}\displaystyle  
\frac{1}{2}\bigg(\frac{Q_1^2}{(h_1^+)^2}-\frac{Q_2^2}{(h_2^+)^2}\bigg) +bh_1^+   
+\bar{b}\eta^+=J, \\[4mm]\displaystyle
\frac{Q_1^2}{h_1^+}+ \frac{Q_2^2}{h_2^+}\bigg(1-\frac{H}{2h_2^+}\bigg)+ \frac{Q_m^2}{\eta^+}+ \frac{J_s^2}{Q_m^2}(\eta^+)^3+ \frac{b}{2}(h_1^+)^2+ \bigg(h_1^+ +\frac{\eta^+}{2} \bigg) \bar{b}\eta^+ =W.
\end{array} 
\end{equation}
The quantities  $Q_i$, $Q_m$, $J_s$, $J$ and $W$  are continuous across the  jump, so  we know them {\it a priori}. 
Thus, the construction of a discontinuous flow is reduced to finding a solution to algebraic equations (\ref{eq:R-Hugoniot-3}) corresponding to a subcritical flow.

Instead of the conservation law for  $q$ one can also use  the energy equation (\ref{eq:energy-bouss}).  For this,  we can eliminate the pressure $p^*$ from equation (\ref{eq:energy-bouss}) by subtracting the fifth equation in (\ref{eq:CL-3l-Boussinesq}) multiplied by $\bar{Q}$. In this case, instead of $J_s$, the variable 
\[  M=\frac{u_1^3 h_1}{2}+ \frac{(\bar{u}^2+3q^2)\bar{u}\eta}{2}+ \frac{u_2^2(u_2 h_2- \bar{Q})}{2}+(bh_1+\bar{b}\eta)u_1 h_1+ \bar{u}\bar{b}\eta(h_1+\eta) \]	
conserves across  the  jump. By combining $W$ and $M$ which are also conserved across the jump, we can eliminate $q$ and obtain  two algebraic equations for $h_1^+$ and $\eta^+$. In this case, the second equation in (\ref{eq:R-Hugoniot-3}) should be replaced by
\begin{equation} \label{eq:R-Hugoniot-add}  
	\begin{array}{l}\displaystyle  
	\frac{1}{2}\bigg(\frac{3Q_m}{\eta^+}-\frac{Q_1}{h_1^+}\bigg)\frac{Q_1^2}{h_1^+} +\frac{Q_m^3}{(\eta^+)^2} +\bigg(\frac{3Q_m}{\eta^+} \Big(\frac{H}{2}-h_1^+ -\eta^+\Big)-Q_2+\bar{Q}\bigg)\frac{Q_2^2}{2(h_2^+)^2}   \\[3mm]\displaystyle
	\quad\quad +\frac{3b Q_m (h_1^+)^2}{4\eta^+} -(bh_1^++\bar{b}\eta^+)Q_1+ \frac{Q_m\bar{b}}{2}\Big(h_1^+-\frac{\eta^+}{2}\Big) =\frac{3Q_m}{2\eta^+}W-M.
	\end{array} 
\end{equation}
If $h_1^+$ and $\eta^+$ are defined, then the velocities behind the jump are found from the first three relations (\ref{eq:vel-plus}). Then we find $q^+$ using the continuity of $W$ across  the jump, since $J_s$ is not conserved in this case.

\begin{figure}
	\begin{center}
		\resizebox{.95\textwidth}{!}{\includegraphics{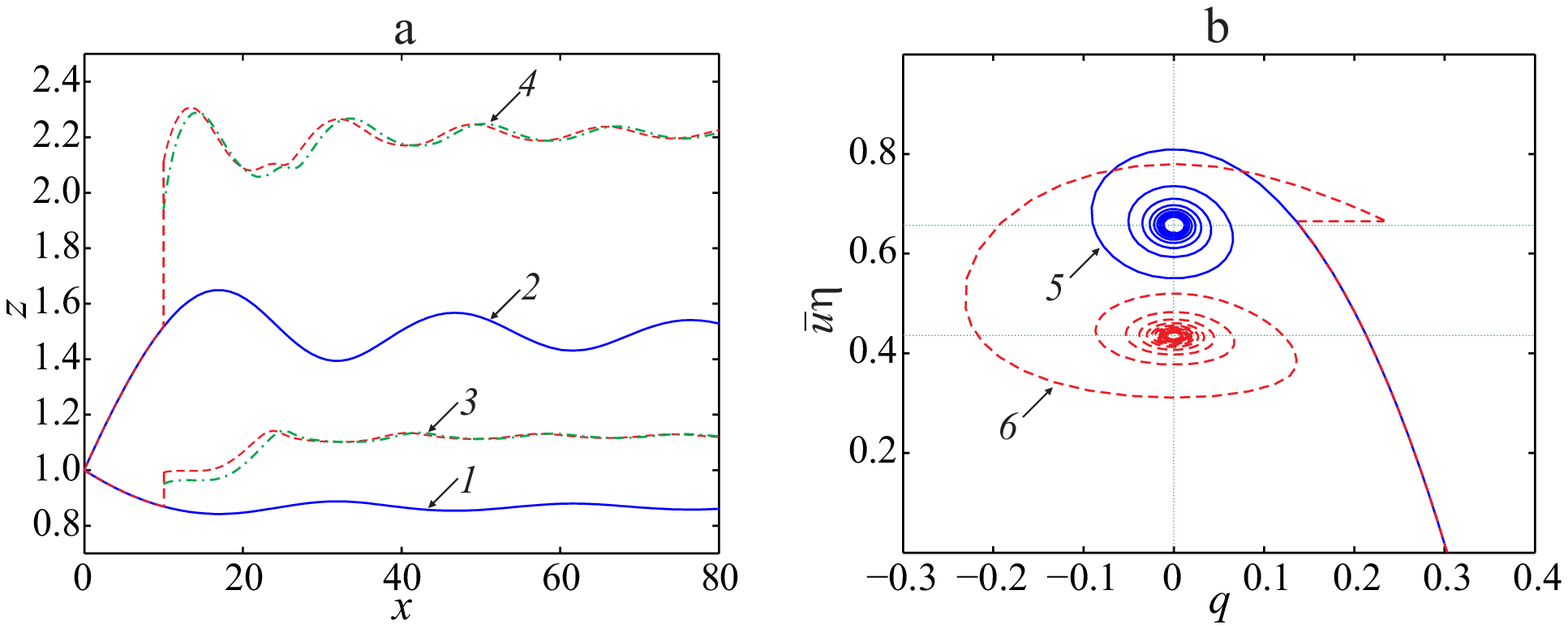}}\\[0pt]
		{\caption{Stationary mixing layer: a --- solid curves {\it 1}, {\it 2} are interfaces $z=h_1$ and $z=h_1+\eta$ for continuous solution of equations (\ref{eq:st-3l-norm}), dashed and  dash-dotted curves {\it 3} and {\it 4} for discontinuous solution with a jump at $x=10$ (the dashed lines correspond to the conservation of  $\bar{u}q$, while dash-dotted lines  correspond to the conservation of energy); b --- solid curve {\it 5} and dashed curve {\it 6} correspond to the continuous and discontinuous solutions on the $(q,\bar{u}\eta)$-plane calculated on the interval $x\in (0,300)$.} \label{fig:fig_2}} 
	\end{center}
\end{figure}

Let us consider the formation of a stationary mixing layer in a supercritical flow over flat bottom ($Z=0$). All variables  considered below are dimensionless. We take $\kappa=6$,  $H_0=10$ and $b=1$.  At   $x=0$ we take $u_{10}=1.8$, $u_{20}=0.6$, $h_{10}=1$, $\eta_0=0.005$ (to carry out computations, it is necessary to set a small but positive value of the intermediate layer thickness). According to formulas (\ref{eq:u0}) and (\ref{eq:q0}) at $x=0$ we obtain $\bar{u}_0=1.118$, $q_0=0.303$ (the second  root of equation (\ref{eq:u0}) $\bar{u}_0=0.322$ lies outside the interval $(u_{10},u_{20})$ and therefore this root is not physically admissible). In virtue of the third equation~(\ref{eq:st-3l-norm}) the buoyancy $\bar{b}_0=b/2=0.5$ remains constant for all $x>0$. It is easy to verify that the data prescribed at $x=0$ correspond to supercritical flow, since all roots $\lambda=\lambda_i$ of characteristic equation~(\ref{eq:chi}) are positive. The corresponding continuous solution of equations (\ref{eq:st-3l-norm}) (interfaces $z=h_1$ and $z=h_1+\eta$) are shown in Fig.~\ref{fig:fig_2}\,a by solid lines. As can be seen from the graph, the monotonic expansion of the mixing layer takes place up to $x\approx 14$. Then its quasi-periodic contraction and expansion occurs. This behaviour is determined by a change in the sign of the variable $q$ (`shear velocity'), which is responsible for the entrainment of liquid into the intermediate mixing layer. The oscillatory nature of the solution is illustrated in Fig.~\ref{fig:fig_2}\,b, which shows the considered solution in the plane $(q,\bar{u}\eta)$ (solid curve). We note that for stationary solutions the extrema of the flow rate $\bar{u}\eta$ in the mixing layer (points of maximum and minimum fluid entrainment) correspond to  $q=0$. When $x\to\infty$, the solution tends towards equilibrium. This flow is supercritical everywhere.

As noted above, it is possible to construct a discontinuous solution. Suppose that in the previous supercritical flow at some point $x=x_s$ there is a shock. Let us choose $x_s=10$. Using the known values of the solution before at $x_x-0$ ($h_1^-=0.869$, $\eta^-=0.649$, $u_1^-=1.689$, $\bar{u}^-=1.028$, $q^-=0.135$) we solve equations (\ref{eq:R-Hugoniot-3}) and find the subcritical state behind the shock ($h_1^+=0.991$, $\eta^+=1.121$, $u_1^+=1.490$, $\bar{u}^+=0.594$, $q^+=0.233$). The choice of this solution comes from the analysis of the Rankine--Hugoniot relations~(\ref{eq:R-Hugoniot-3}) and will be discussed  later. Then we solve ODEs (\ref{eq:st-3l-norm}) with these data at $x=x_s$. The obtained discontinuous solution is shown in Fig.~\ref{fig:fig_2} by dashed curves. This flow is subcritical in the region $x>x_s$ and also has an oscillatory character. 

Nonlinear algebraic system (\ref{eq:R-Hugoniot-3}) in the considered case has four admissible solutions (such that $h_1^+>0$, $\eta^+>0$, $h_1^+ +\eta^+<H$ and $(h_1^+, \eta^+) \neq (h_1^-, \eta^-)$). One of them ($h_1^+=0.561$, $\eta^+=7.912$) transforms the supercritical flow into the supercritical one. For another solution ($h_1^+=8.512$, $\eta^+=0.037$) the characteristic equation~(\ref{eq:chi}) has complex roots. Both solutions are likely to be unstable and are not realized. Finally, the last  admissible solutions of equations~(\ref{eq:R-Hugoniot-3}) ($h_1^+=0.991$, $\eta^+=1.121$, and $h_1^+=3.051$, $\eta^+=4.946$) correspond  to the subcritical flow behind the shock.  We chose the  solution having the  shock of smallest amplitude. To justify this,  we will use now  non-stationary equations (\ref{eq:model}) and will construct a numerical discontinuous solution arising in the case of flow over an obstacle. By choosing the height of the obstacle, it is possible to achieve a quasi-stationary regime in which the solution $h_1^+=0.991$ and  $\eta^+=1.121$ is realized. 

\begin{figure}
	\begin{center}
		\resizebox{.95\textwidth}{!}{\includegraphics{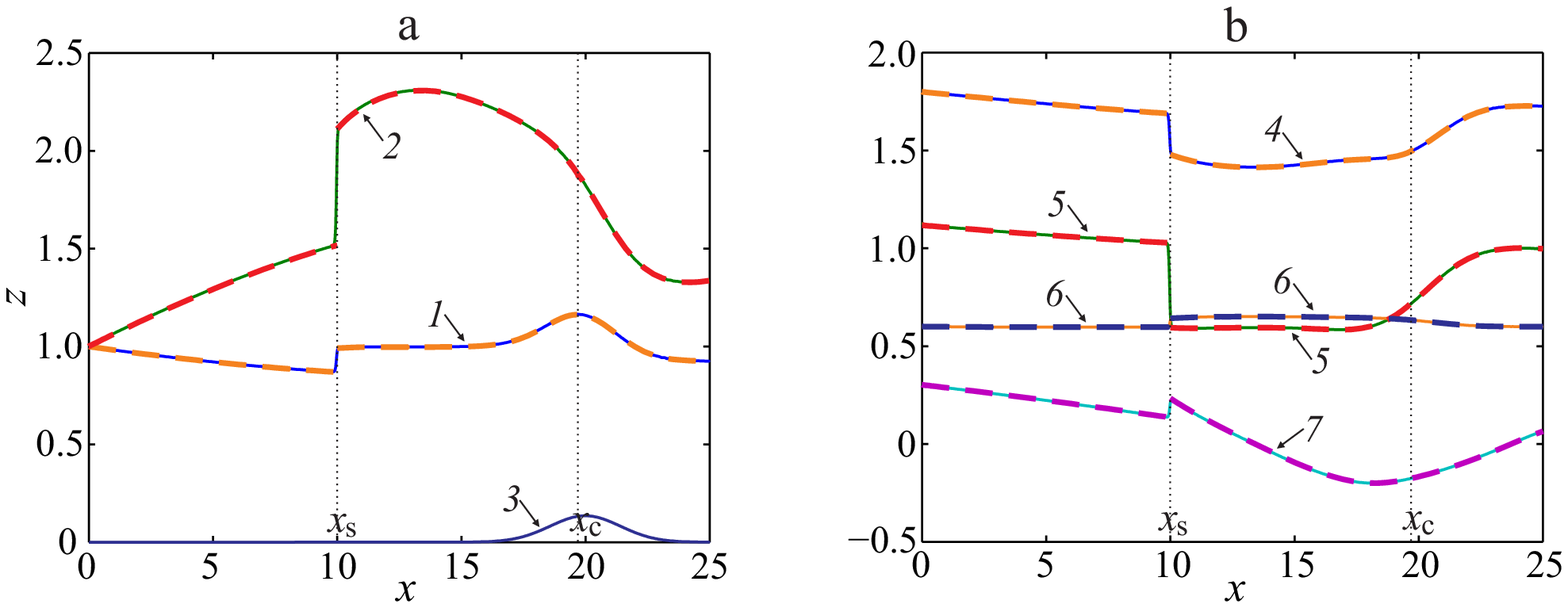}}\\[0pt]
		{\caption{Quasi-stationary discontinuous flow: a --- interfaces $z=Z+h_1$ and $z=H_0-h_2$ (curves {\it 1}, {\it 2}) and bottom topography $z=Z(x)$ (curve {\it 3}); b --- velocities $u_1$, $\bar{u}$ and $u_2$ in the layers (curves {\it 4}, {\it 5} and {\it 6}) and variable $q$ (curve {\it 7}). Solid curves --- solution of equations (\ref{eq:model}) at $t=400$, dashed curves --- corresponding stationary solution.} \label{fig:fig_3}} 
	\end{center}
\end{figure}

It is interesting to note that the use of the energy equation across the jump, i.e. of the system consisting of the first equation of (\ref{eq:R-Hugoniot-3}) and (\ref{eq:R-Hugoniot-add}) under the same conditions ahead of the jump gives only two admissible solutions $h_1^+=8.526$, $\eta^+=0.025$ and $h_1^+=0.950$, $\eta^+=0.991$. The first of these solutions gives the complex value of the variable $q$ and therefore is not physical. The second solution corresponds to $u_1^+=1.544$, $\bar{u}^+=0.672$ and $q=0.302$. All roots of the characteristic polynomial (\ref{eq:chi}) are real. One of the roots is negative, the other roots are positive. This corresponds to the subcritical flow regime behind the jump. The corresponding downstream solution  (for $x>x_s$) is shown in Fig.~\ref{fig:fig_2} by a dash-dotted curve. It can be clearly seen that  replacing the cumbersome energy equation by its differential consequence does not lead to a significant difference in the solution  behaviour.

To conclude this Section, let us consider the  formation of a quasi-stationary flow with a jump over an obstacle. As before, we take $H_0=10$, $b=1$ and set the same data $u_{10}=1.8$, $u_{20}=0.6$, $h_{10}=1$, $h_{20}=8.995$ $\eta_0=0.005$, $\bar{u}_0=1.118$, $q_0=0.303$ and $\bar{b}_0=0.5$  on the left boundary of the computational domain $x=0$. The initial conditions for unknowns are the same excepting  $h_1$ for which  one takes  $h_1(x,0)=h_{10}-Z(x)$, $x>0$.   We set `soft' conditions $\bU_N=\bU_{N-1}$ on the right boundary of the computational domain $x=25$. Here $\bU_j$ is the value of unknown vector-function $\bU$ in the nodal point $x_j$. The bottom is chosen as
\[ Z(x)=z_0\exp(-0.25(x-x_0)^2) \]
with $x_0=20$ and $z_0=0.135$. The bottom is required to create  a left-facing shock. The height of the obstacle $z_0$ is selected in such a way  that the shock propagates slowly or stops at some point $x=x_s$. This corresponds to the formation of a stationary discontinuous solution. To solve balance laws (\ref{eq:model}) numerically, we implemented the second-order central scheme proposed by \cite{NT90}. The calculation results on a uniform grid with the number of nodes $N=1000$ on the interval $x\in (0,10)$ at $t=400$ are shown in Fig.~\ref{fig:fig_3} by solid curves. The position of the shock front is at $x=10$. This shock solution corresponds to our stationary solution having the smallest  amplitude. 

The corresponding discontinuous stationary solution of equations (\ref{eq:st-3l-norm}) is shown in Fig.~\ref{fig:fig_3} by dashed lines. Visually, there is no  difference between the stationary solution and that  obtained from non-stationary computations. We note that the  discontinuous solutions shown in   Fig.~\ref{fig:fig_2} and ~\ref{fig:fig_3}) coincide, further difference  is related to  the bottom topography. The flow shown in Fig.~\ref{fig:fig_3} continuously passes from subcritical regime to supercritical one at the point $x_c\approx 19.7$. Thus, at the end of  the computational domain the flow again becomes supercritical.

\section{Model validation}
\label{sec:model_validation}

In this section we compare  numerical simulation and experimental or field data found in the literature.  In subsections \ref{transcritical} and \ref{evolution_supercritical} we use the SGS system of units  (cm, g, s) while in \ref{evolution_subcritical} the International System of units is used (m, kg, s). 

\subsection{Transcritical flow over an obstacle}
\label{transcritical}

We consider the formation of the mixing layer in a  down-flow current of a dense fluid  in the framework of the proposed three-layer model.  The formulation of the problem is close to \cite{Pawlak00}, where mixing and entrainment during the evolution of stratified flows were experimentally studied.

\begin{figure}
	\begin{center}
		\resizebox{.95\textwidth}{!}{\includegraphics{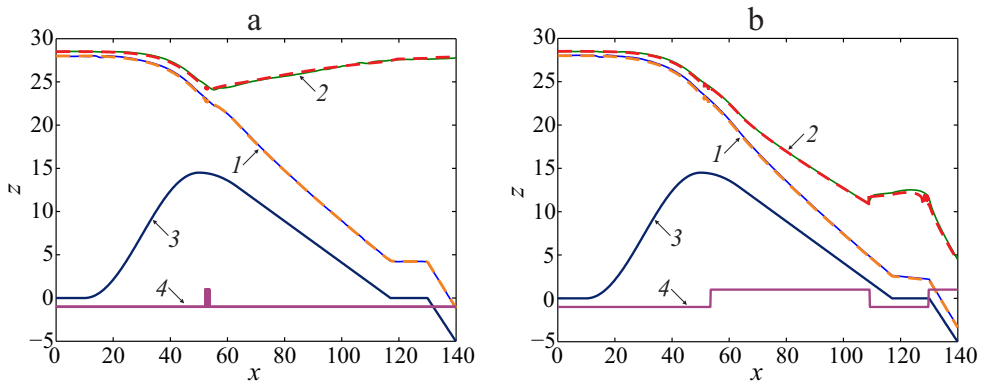}}\\[0pt]
		{\caption{Flow over an obstacle: a --- subcritical regime; b --- supercritical flow on the leeward side of the obstacle (curve {\it 3}). Solid curves {\it 1} and {\it 2} show the interfaces $z=Z+h_1$ and $z=H_0-h_2$ at $t=t_*$ obtained by non-stationary model (\ref{eq:model}), line {\it 4} presents the sign of the characteristic function $\chi(0)$. Dashed curves --- corresponding stationary solution of equations (\ref{eq:st-3l-norm}).} \label{fig:fig_PA}} 
	\end{center}
\end{figure}

\begin{figure}
	\begin{center}
		\resizebox{.475\textwidth}{!}{\includegraphics{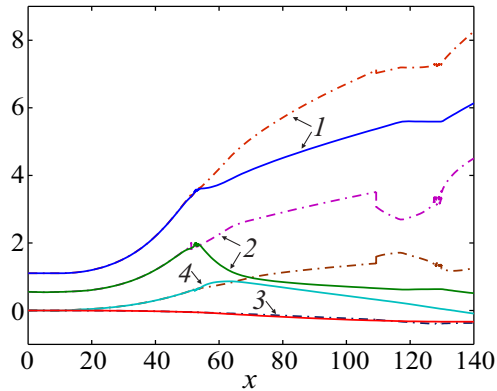}}\\[0pt]
		{\caption{Velocities in a transcritical stationary flow. Solid curves correspond to the flow regime shown in Fig.~\ref{fig:fig_PA}\,a, dash-dotted lines --- Fig.~\ref{fig:fig_PA}\,b. Curves {\it 1, 2} and {\it 3} are the velocities in the lower, middle and upper layers; curve {\it 4} shows the variable $q$.} \label{fig:fig_PA_vel}} 
	\end{center}
\end{figure}

In this test we take $\kappa=6$ and choose $H_0=50$, $b=1.2$. Computations are carried out in the domain $x\in [0,L]$, $L=140$. The bottom topography is shown in Fig.~\ref{fig:fig_PA} (curve {\it 3}).  The obstacle is located on the interval $x\in (10,117.5)$ and its maximum height $z_{\max}=14.5$ is reached at the point $x=50$. The leeward side of the obstacle is rectilinear with a slope of $k=-0.24$ (approximately $13.5^\circ$). Near the outlet section of the channel ($x>130$), the bottom has a slope of $k=-0.5$. On the left boundary $x=0$ we set $h_1=28$, $\eta=0.5$, $u_1=U$, $\bar{u}=U/2$, $u_2=0$, $\bar{b}=b/2$ and $q=0$. First, we construct stationary solutions for $U\approx 1.104$.
At such a velocity, the flow becomes critical over the obstacle since  $\chi(0)$ vanishes at $x\approx 53.6$ (see the definition \eqref{eq:chi} of the characteristic polynomial $\chi(\lambda)$). Fig.~\ref{fig:fig_PA}\,a (dashed curves) is obtained for $U=1.1039$ by using the standard {\sf ode45} procedure of the MATLAB package that implements the fourth-order Runge–Kutta method. As can be seen from the graph, intensive mixing occurs on the leeward side of the obstacle, and fluid from the outer layers is entrained into the mixing layer. The fluid velocities in the layers and the variable $q$, which is responsible for the mixing process, are shown in Fig.~\ref{fig:fig_PA_vel} (solid curves). For $x<x_c$ ($x_c$ is a critical point near the top of the obstacle, where $\chi(0)$ vanishes), the flow velocity in the lower and intermediate layers increases. For $x>x_c$, the heavy fluid continues to accelerate along the leeward slope, while the fluid of intermediate density in the mixing layer, on the contrary, is decelerated. The velocity of the light fluid in the upper layer changes insignificantly, decreasing monotonically over the entire interval. In this case, the solution is subcritical everywhere, with the exception of a small area above the obstacle. 

A slight increase of the lower layer velocity leads to a significant change of the flow for $x>x_c$. Fig.~\ref{fig:fig_PA}\,b (dashed curves) is obtained for $U=1.1052$. With this velocity, the flow over the leeward side of the obstacle is supercritical and the mixing layer thickness is noticeably smaller than in the case of subcritical flow. Both the heavier fluid in the lower layer and the lighter fluid in the intermediate mixing layer are accelerated over the leeward slope of the obstacle (Fig.~\ref{fig:fig_PA_vel}, dash-dotted curves). Before reaching the  flat bottom, the transition  from supercritical to subcritical flow occurs through a hydraulic jump at $x\approx 109.35$ (this point is chosen on the basis of a non-stationary computation). In the subcritical flow region, the mixing layer thickness increases, and the fluid velocity in the intermediate layer slows down. Further, in the vicinity of the outlet section, the flow again becomes  supercritical. We note that by virtue of the third equation (\ref{eq:st-3l-norm}) and the condition $\bar{b}=b/2$ at $x=0$, one obtains $\bar{b}=b/2$ in the entire flow region.

The constructed stationary solutions can also be obtained as a result of the numerical solution of non-stationary equations (\ref{eq:model}). The computations  are carried out on a uniform grid with a number of nodes $N=1000$. As the initial data at $t=0$, we take the above-mentioned values of the functions at $x=0$, with the exception of $h_1$ and $u_1$, which are defined as $h_1=h_{10}-Z(x)$, $u_1=(Q-\bar{u}_0\eta_0)/h_1$. Here $Q=(h_{10}+\eta_0/2)U$ and index `0' corresponds to the values of functions at $x=0$. The `soft' conditions $\bU_N=\bU_{N-1}$ are imposed on the right boundary $x=L$ of the computational domain. Numerical solutions of equations (\ref{eq:model}) obtained for $U=1.10$  and $U=1.11$  are shown in Fig.~\ref{fig:fig_PA}\,a  and Fig.~\ref{fig:fig_PA}\,b, respectively. Both graphs are shown at $t=500$\,s, which corresponds to reaching the stationary regime. As can be seen from the figure, there is good agreement between stationary and non-stationary computations, both for the subcritical and supercritical regimes of flow over the leeward side of the obstacle. Note that in non-stationary computations, the threshold value of the velocity $U$ can vary slightly depending on the chosen numerical scheme and spatial resolution.

\subsection{Evolution of a mixing layer in a supercritical flow: comparison with experiment}
\label{evolution_supercritical}

In this section we  consider the evolution of the mixing layer flowing down  the inclined bottom and compare solutions of  stationary equations~(\ref{eq:st-3l-norm}) with the experimental data by ~\cite{Pawlak00}. In these experiments the channel height was  $H_0=30$ and a constant bottom slope  was  $10.8^\circ$. At  $x=x_0=10.5$  the depths of the layers and corresponding average velocities are as follows :  $h_1=6.25$, $\eta_0=0.75$, $h_2=8$ and $u_1=4$, $\bar{u}=u_1/2$, $u_2=0$. We also choose $b=1.4$, $\bar{b}=b/2$. This is consistent with the flow parameters in \cite{Pawlak00} (Fig.~9 therein). Finally, $q$ for $x=x_0$ can take  arbitrary value  from the interval $[0,1]$. This has no noticeable effect on the position of  the interfaces $z=Z+h$ and $z=Z+h+\eta$. To be specific, we take $q=0$. It is easy to verify that such a choice of flow parameters at $x=x_0$ corresponds to the supercritical regime, since $\chi(0)>0$. The dissipation parameter is taken as $\kappa=8$. 

\begin{figure}
	\begin{center}
		\resizebox{.75\textwidth}{!}{\includegraphics{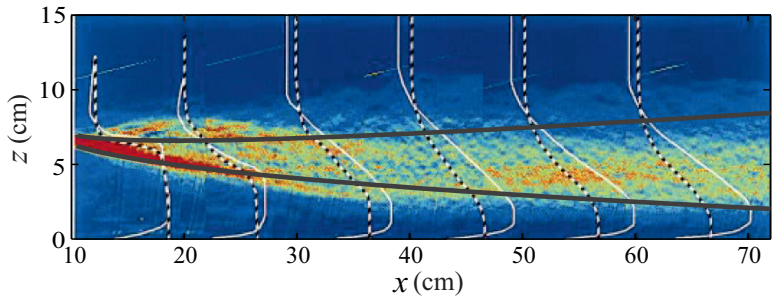}}\\[0pt]
		{\caption{The interfaces $z=Z+h$ and $z=Z+h+\eta$ according to equations (\ref{eq:st-3l-norm}) are shown by thick solid curves. Color  picture presents experimental data of the density gradient field images with velocity and density profiles (thin transverse curves) \cite{Pawlak00} (Fig.~9 therein). The picture is rotated so that the slope becomes horizontal.} \label{fig:fig_PA_exp}} 
	\end{center}
\end{figure}

In Fig.~\ref{fig:fig_PA_exp} bold solid lines show the boundaries $z=h_1$  and $z=h_1+\eta$ of the mixing layer in the interval $x \in (10.5, 72)$ determined by model~(\ref{eq:st-3l-norm}). The constant slope bottom has been rotated  to become horizontal,   and the abscissa  on Fig.~\ref{fig:fig_PA_exp} is the downslope distance. Coloured picture represents the density gradient field image (blue is for  a  lower density gradient, red is for  a higher one) obtained by \cite{Pawlak00}. As we can see, there is a fairly good agreement between the experimental data and our numerical solution. We have  to mention that in the experiment (as well as in the nature), there is a slight backward flow above the mixing layer. This fact is reflected in our model. In general, the nature of the flow is similar to the example considered above in the case of a supercritical flow over the leeward side of the obstacle (see Fig.~\ref{fig:fig_PA}\,b and \ref{fig:fig_PA_vel}). 

Note that a similar comparison of the mixing layer boundaries with the experimental data of \cite{Pawlak00} was carried out by \cite{LiapDG18} (see Fig.~2 in this work) based on a different (albeit similar) depth-averaged model. The constitutive equations proposed by \cite{LiapDG18} better describe the mixing layer for $x\in (10,25)$, but give excessively overestimated sizes of the mixing region for $x>30$. Our proposed model gives better results for a sufficiently developed mixing layer.

\subsection{Evolution of a mixing layer in a subcritical flow: comparison with field observations}
\label{evolution_subcritical}

Field observations  of  stratified flows over topography  in Knight Inlet (British Columbia, Canada) are given by \cite{Farmer99, Cummins06}. The stratified fluid flow  over   a sill illustrates the upstream formation of a strong internal bore. In particular, a flow regime was recorded over the leeward side of the obstacle similar to that shown in Fig.~\ref{fig:fig_PA}\,a. An image of acoustic backscatter obtained with a 200-kHz echosounder is presented by \cite{Cummins06} (see in Fig.~3 therein), where blue and red colours correspond to 30 dB and 55 dB, respectively. This shows a plunging flow on the lee side of the sill with large-amplitude instabilities on the interface of the downslope flow. The Kelvin--Helmholtz instability develops as a result of the large shear found along the interfaces. The formation of the flow structure shown in the figure has been thoroughly documented by \cite{Farmer99}.

\begin{figure}
	\begin{center}
		\resizebox{.75\textwidth}{!}{\includegraphics{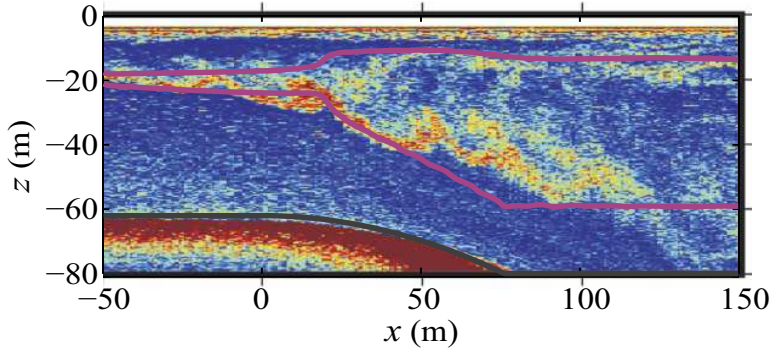}}\\[0pt]
		{\caption{The interfaces $z=Z+h$ and $z=Z+h+\eta$ according to model (\ref{eq:model}) are shown by bold solid curves. Coloured picture presents field data of acoustic backscatter obtained by \cite{Cummins06} (Fig.~3 therein).} \label{fig:fig_CAV}} 
	\end{center}
\end{figure}

We do not claim here to accurately reproduce the field data. Nevertheless, choosing the flow parameters close to those considered by \cite{Cummins06}, our model describes the characteristic features of the mixing layer evolution. We perform computations  using three-layer equations  (\ref{eq:model}) in the domain $x\in [-50, 150]$ (in meters). The upper fluid level is constant and equal to zero. On the interval $x\in [-50, 0]$, the total depth $H_0-Z(x)$ is $62$, and then, according to the parabolic law, increases to $80$ for $x\in [0, 72]$ and then takes this constant value. Buoyancy $b$ is $0.06$. On the left boundary $x=-50$ of the calculated interval, the layers thicknesses, corresponding velocities and buoyancy  are set as follows $h_1=41$, $\eta=3.5$ and $h_2=17.5$; $u_1=1$, $\bar{u}=0.59$ and $u_2=0.18$, $\bar{b}=b/2$, $q=u_2$. The same values of the functions are chosen as the initial conditions, excepting those for  $h_1$ and $u_1$. The last  are determined taking into account the bottom topography and the constancy of the total flow rate. As before, `soft' boundary conditions are imposed at $x=150$.  The dissipation constant is taken as $\kappa=8$. Over time, the numerical solution reaches a quasi-stationary regime. In fig.~\ref{fig:fig_CAV}, bold solid lines show the interface between the layer obtained according to model (\ref{eq:model}) at $t=4000$. The coloured image on this figure shows the acoustic backscatter field data \cite{Cummins06} (Fig.~3 therein). As can be seen from this figure, the interfaces predicted by our model are in a good agreement with the field data of the acoustic backscatter \cite{Cummins06}, where the interfaces correspond to the strongest acoustic response. We note that the obtained numerical solution is subcritical $\chi(0)<0$ in all computational domain. Obviously, this solution is similar to that considered above in the case of a subcritical flow on the leeward side of the obstacle (see Fig.~\ref{fig:fig_PA}\,a and  Fig.~5).

\section{Conclusion}
\label{sec:conclusion}

A long wave approximation of  the  non-homogeneous Euler equations was used for the construction of  a depth-averaged model of  three layer flows. The outer layers are homogeneous and almost potential and can be described by Saint-Venant-like equations (\ref{eq:upper-layer}) and (\ref{eq:lower-layer}). The fluid flow in the intermediate layer is sheared and inhomogeneous, and is  described by equations (\ref{eq:ML-2}). The interfaces separating these layers are considered as fronts through which the turbulent mixing of fluids occurs. The general  model (\ref{eq:CL-3l}) reveals the main mechanisms of the mixing layer development. The model is reduced to the classical equations of three-layer shallow water with a sheared  intermediate layer if the  mass transfer between the layers is absent.

In the Boussinesq approximation, the proposed model is simplified and takes the form (\ref{eq:model}), which allows a simple numerical implementation.  It was proved  that for small relative velocities  in the layers, the model is hyperbolic. The concepts of supercritical, subcritical and transcritical flows are defined in terms of the signs of the eigenvalues of the corresponding characteristic polynomial. It makes it possible to formulate in classical terms the conditions for the control of a subcritical flow by a downstream located obstacle. The system of conservation laws representing our  model uniquely determines the jump relations. It is established that the use of an additional conservation law for the shear velocity  instead of the energy equation simplifies the study of discontinuous solutions and  does not lead to a significant change in their structure. The model is able to control  the mixing process by changing the position of the stationary hydraulic jump upstream of the obstacle. For supercritical and subcritical co--current flows  new oscillating structure of the mixing layer is found.

Considerable attention is paid to the validation  of the proposed model (\ref{eq:model}) by comparing its solutions with  experimental data and field observations found in the literature. We study transcritical three-layer flows over an obstacle. This  problem is close to that studied in \cite{Pawlak00}, where mixing and entrainment during the evolution of stratified flows were experimentally explored. It is shown that the proposed model describes quite accurately the development of the mixing layer on the leeward side of the obstacle in the supercritical flow regime (Fig.~\ref{fig:fig_PA_exp}). Field observations of the interaction of a stratified flow with topography presented by \cite{Farmer99, Cummins06} were also used to validate the model. Comparison of the numerical  results with the observations  over the   Knight Inlet sill illustrating the formation of a mixing layer, is shown in Fig.~\ref{fig:fig_CAV}. These results demonstrate the  ability of our model to describe the characteristic features of gravity currents.

\section*{Acknowledgements}
A.C. and V.L. were supported by the Russian Science Foundation (grant No.\,21-71-20039).  S.G. has been partially funded by the Excellence Initiative of Aix-Marseille UniversityA*Midex, a French Investissements d’Avenir programme AMX-19-IET-010.

%\backsection[Acknowledgements] {A.C. and V.L. were supported by the Russian Science Foundation (grant No.\,21-71-20039).  S.G. has been partially funded by the Excellence Initiative of Aix-Marseille UniversityA*Midex, a French Investissements d’Avenir programme AMX-19-IET-010.}

%\backsection[Declaration of interests]{The authors report no conflict of interest.}

%\backsection[Author ORCID]{A. Chesnokov, https://orcid.org/0000-0003-0959-8806; S. Gavrilyuk, https://orcid.org/0000-0003-4605-8104; V. Liapidevskii, https://orcid.org/0000-0003-4250-8497}

%\backsection[Author contributions]{All authors contributed equally.}

%\bibliographystyle{jfm}


\begin{thebibliography}{99}

\bibitem[Armi \& Farmer (2002)]{Armi02}
{\sc Army, L. \& Farmer, D.} 2002 Stratified flow over topography: bifurcation fronts and transition to the uncontrolled state. {\it Proc. R. Soc. Lond. A} {\bf 458}, 513--538. %doi: 10.1098/rspa.2001.0887

\bibitem[Baines (2005)]{Baines05}
{\sc Baines, P.G.} 2005 Topographic Effects in Stratified Flows. Cambridge University Press.

\bibitem[Baines (2016)]{Baines16}
{\sc Baines, P.G.} 2016 Internal hydraulic jumps in two-layer systems. {\it J. Fluid Mech.} {\bf 787}, 1--15. %doi: 10.1017/jfm.2015.662

\bibitem[Barros {\it et al.} (2007)]{Barros07}
{\sc  Barros, R., Gavrilyuk, S. \& Teshukov, V.} 2007 Dispersive nonlinear waves in two-layer flows with free surface. I. Model derivation and general properties. {\it Stud. Appl. Math.} {\bf 119}, 191--211.

\bibitem[Chesnokov \& Liapidevskii (2020)]{ChesnLiap20}
{\sc Chesnokov, A.A. \& Liapidevskii, V.Yu.} 2020 Mixing layer and turbulent jet flow in a Hele--Shaw cell. {\it Int. J. Non-Linear. Mech.} {\bf 125}, 103534. %doi: 10.1016/j.ijnonlinmec.2020.103534

\bibitem[Chu \& Baddour (1984)]{Chu84}
{\sc  Chu, V.H. \& Baddour, R.} 1984 Turbulent gravity-stratified shear flows. {\it J . Fluid Mech.} {\bf 138}, 353--378. %doi: 10.1017/S002211208400015X

\bibitem[Cummins {\it et al.} (2006)]{Cummins06}
{\sc Cummins, P.F., Armi,  L. \& Vagle, S.} 2006 Upstream internal hydraulic jumps. {\it J. Phys. Oceanogr.} {\bf 36}, 753--769.

\bibitem[Farmer \& Armi (1999)]{Farmer99}
{\sc Farmer, D. \& Armi, L.} 1999 Stratified flow over topography: The role of small-scale entrainment and mixing in flow establishment. {\it Proc. R. Soc. Lond. A} {\bf 455}, 3221--3258. %doi: 10.1098/rspa.1999.0448

\bibitem[Gavrilyuk {\it et al.} (2016)]{GLCh16}
{\sc Gavrilyuk, S.L., Liapidevskii, V.Yu. \& Chesnokov, A.A.} 2016 Spilling breakers in shallow water: applications to Favre waves and to the shoaling and breaking of solitary waves. {\it J. Fluid Mech.} {\bf 808}, 441--468. %doi:10.1017/jfm.2016.662

\bibitem[Gavrilyuk {\it et al.} (2019)]{GLCh19}
{\sc Gavrilyuk, S.L., Liapidevskii, V.Yu. \& Chesnokov, A.A.} 2019 Interaction of a subsurface bubble layer with long internal waves. {\it Eur. J. Mech. B Fluids} {\bf 73}, 157--169.  %doi:10.1016/j.euromechflu.2017.07.004

\bibitem[Horsley \& Woods (2018)]{Horsley18}
{\sc Horsley, M.C. \& Woods, A.W.} 2018 A note on analytic solutions for entraining stratified gravity currents. {\it J. Fluid Mech.} {\bf 836}, 260--276. %doi:10.1017/jfm.2017.834

\bibitem[Jagannathan {\it et al.} (2017)]{Jagannathan17}
{\sc Jagannathan, A., Winters, K.B. and Armi, L.} 2017 Stability of stratified downslope flows with an overlying stagnant isolating layer. {\it J. Fluid Mech.} {\bf 810}, 392--411. %doi: 10.1017/jfm.2016.683

\bibitem[Jagannathan {\it et al.} (2020)]{Jagannathan20}
{\sc Jagannathan, A., Winters, K.B. \& Armi, L.} 2020 The effect of a strong density step on blocked stratified flow over topography. {\it J. Fluid Mech.} {\bf 889}, A23. %doi: 10.1017/jfm.2020.87

\bibitem[Lamb (2004)]{Lamb04}
{\sc Lamb, K.G.} 2004 On boundary–-layer separation and internal wave generation at the Knight Inlet sill, {\it Proc. R. Soc. Lond. A. } {\bf 460}, 2305--2337. %10.1098/rspa.2003.1276 

\bibitem[Liapidevskii (2004)]{Liap04}
{\sc Liapidevskii} 2004 Mixing layer on the lee side of an obstacle. {\it J. Appl. Mech. Tech. Phys.} {\bf 45}, 199--203. %doi: 10.1023/B:JAMT.0000017582.70655.d9

\bibitem[Liapidevskii \& Chesnokov (2014)]{LiapChesn14} 
{\sc Liapidevskii, V.Yu. \& Chesnokov, A.A.} 2014 Mixing layer under a free surface. {\it J. Appl. Mech. Tech. Phys.} {\bf 55}, 299--310. %doi:10.1134/S0021894414020126

\bibitem[Liapidevskii {\it et al.} (2018)]{LiapDG18}
{\sc Liapidevskii, V.Yu., Dutykh, D. \& Gisclon, M.} 2018 On the modelling of shallow turbidity flows. {\it Adv. Water Resour.} {\bf 113}, 310--327. %doi: 10.1016/j.advwatres.2018.01.017

\bibitem[Lipatov {\it et al.} (2021)]{LipLiapCh21}
{\sc Lipatov, I.I., Liapidevskii, V.Yu. \& Chesnokov, A.A.} Forced oscillations of a pseudoshock in transonic gas flow in a diffuser, Fluid Dyn. 56 (2021) 860--869. %doi: 10.1134/S0015462821060094

\bibitem[Nessyahu \& Tadmor (1990)]{NT90} 
{\sc Nessyahu, H. \& Tadmor, E.} 1990 Non-oscillatory central differencing schemes for hyperbolic conservation laws. {\it J. Comp. Phys.} {\bf 87}, 408--463. %doi: 10.1016/0021-9991(90)90260-8

\bibitem[Ogden \& Helfrich (2016)]{Ogden16}
{\sc Ogden, K.A. \& Helfrich, K.R.} 2016 Internal hydraulic jumps in two-layer flows with upstream shear. {\it J. Fluid Mech.} {\bf 789}, 64--92. %doi:10.1017/jfm.2015.727

\bibitem[Ogden \& Helfrich (2020)]{Ogden20}
{\sc Ogden, K.A. \& Helfrich, K.R.} 2020 Internal hydraulic jumps in two-layer flows with increasing upstream shear. {\it Phys. Rev. Fluids} {\bf 5}, 074803. %doi:10.1103/PhysRevFluids.5.074803

\bibitem[Pawlak \& Armi (1998)]{Pawlak98}
{\sc Pawlak, G. \& Armi, L.} 1998 Vortex dynamics in a spatially accelerating shear layer. {\it J. Fluid Mech.} {\bf 376}, 1--35. %doi: 10.1017/S002211209800250X

\bibitem[Pawlak \& Armi (2000)]{Pawlak00}
{\sc Pawlak, G. \& Armi, L.} 2000 Mixing and entrainment in developing stratified currents. {\it J. Fluid Mech.} {\bf 424}, 45--73. %doi: 10.1017/S0022112000001877

\bibitem[Pope (2000)]{Pope_2000}
{\sc Pope, S. B.} 2000 Turbulent Flows. Cambridge University Press.

\bibitem[Sher \& Woods (2015)]{Sher15}
{\sc Sher, D. \& Woods, A.W.} 2015 Gravity currents: entrainment, stratification and self-similarity. {\it J. Fluid Mech.} {\bf 784}, 130--162. %doi:10.1017/jfm.2015.576

\bibitem[Sher \& Woods (2017)]{Sher17}
{\sc Sher, D. \& Woods, A.W.} 2017 Mixing in continuous gravity currents. {\it J. Fluid Mech.} {\bf 818}, R4. %doi: 10.1017/jfm.2017.168

\bibitem[Simpson (1997)]{Simpson97}
{\sc  Simpson, J.} 1997 Gravity Currents. Cambridge University Press,.

\bibitem[Teshukov (2007)]{Tesh07}
{\sc Teshukov, V.M.} 2007 Gas-dynamic analogy in the theory of stratified liquid flows with a free boundary. {\it Fluid Dynamics} {\bf 42}, 807--817. %doi: 10.1134/S0015462807050134

\bibitem[Thorpe \& Li (2014)]{Thorpe14}
{\sc Thorpe, S.A. \& Li, L.} Turbulent hydraulic jumps in a stratified shear flow. Part 2. {\it J. Fluid Mech.} {\bf 758}, 94--120.

\bibitem[Thorpe {\it et al.} (2018)]{Thorpe18}
{\sc Thorpe, S.A., Malarkey, J., Voet, G., Alford, M.H. Girton, J.B. \& Carter, G.S.} 2018 Application of a model of internal hydraulic jumps. {\it J. Fluid Mech.} {\bf 834}, 125--148. %doi: 10.1017/jfm.2017.646

%\bibitem[Townsend (1956)]{Townsend56} {\sc Townsend, Townsend} 1956 The structure of turbulent shear flow. Cambridge University Press.

\bibitem[Winters \& Armi (2014)]{Winters14} 
{\sc Winters, K.B. \& Armi, L.} 2014 Topographic control of stratified flows: upstream jets, blocking and isolating layers. {\it J. Fluid Mech.} {\bf 753}, 80--103. %doi: 10.1017/jfm.2014.363

\bibitem[Yuan \& Horner-Devine (2017)]{Yuan17}
{\sc Yuan, Y. \& Horner-Devine, A.R.} Experimental investigation of large-scale vortices in a freely spreading gravity current. {\it Phys. Fluids} {\bf 29}, 106603 %doi: 10.1063/1.5006176

\end{thebibliography}
\end{document}